 \documentclass[aps,pre,twocolumn,showpacs,amsmath,amssymb,superscriptaddress]{revtex4-1}
\usepackage{graphicx}
\usepackage{dcolumn}
\usepackage{bm}
\usepackage{natbib}
\usepackage{amsmath}
\usepackage{array}
\usepackage{hyperref}
\usepackage{cleveref}[2012/02/15]% v0.18.4; 
\crefformat{footnote}{#2\footnotemark[#1]#3}

\begin{document}
{ \title{Universal Scaling in the Aging of the Strong Glass Former
    SiO$_2$}

\author{Katharina Vollmayr-Lee}
 \email{kvollmay@bucknell.edu}
\affiliation{Department of Physics and Astronomy, Bucknell University,
      Lewisburg, Pennsylvania 17837, USA}
%\affiliation{Georg-August-Universit\"at G\"ottingen,
%Institut f\"ur Theoretische Physik,
%Friedrich-Hund-Platz 1, 37077 G\"ottingen, Germany}
\author{Christopher H.~Gorman}
%\affiliation{Department of Physics, Wabash College, Crawfordsville, Indiana 47933, USA}
\affiliation{Department of Mathematics, University of California, Santa Barbara, CA 93106, USA}
\author{Horacio E.~Castillo}
\affiliation{Department of Physics and Astronomy and Nanoscale and
   Quantum Phenomena Institute, Ohio University, Athens, Ohio, 45701,
   USA}

%\date{May 20, 2013}
\date{\today}
\begin{abstract}

We show that the aging dynamics of a strong glass former displays a
strikingly simple scaling behavior, connecting the average dynamics
with its fluctuations, namely the dynamical heterogeneities. 
We perform molecular dynamics simulations of SiO$_2$ with BKS
interactions, quenching the system from high to low temperature, and
study the evolution of the system as a function of the waiting time
$t_{\rm w}$ measured from the instant of the quench. We find that both
the aging behavior of the dynamic susceptibility $\chi_4$ and the
aging behavior of the probability distribution $P(f_{{\rm s},{\mathbf
    r}})$ of the local incoherent intermediate scattering function
$f_{{\rm s},{\mathbf r}}$ can be described by simple scaling forms in
terms of the global incoherent intermediate scattering function
$C$. The scaling forms are the same that have been found to describe
the aging of several fragile glass formers and that, in the case of
$P(f_{{\rm s},{\mathbf r}})$, have been also predicted
theoretically. A thorough study of the length scales involved
highlights the importance of intermediate length scales. We also
analyze directly the scaling dependence on particle type and on
wavevector $q$, and find that both the average and the fluctuations
of the slow aging dynamics are controlled by a unique aging clock,
which is not only independent of the wavevector $q$, but is the same
for O and Si atoms.
% Since SiO$_2$ is known to show drastically different dynamics for Si
% and O atoms at short time scales, we investigate the influence of
% the particle type $\alpha \in \{{\rm Si,O,\textrm{all}}\}$. Our
% results clearly indicate that despite the differences in the
% short-time dynamics, there is a common clock for the slow dynamics
% of Si and O atoms. 

\end{abstract}
%http://www.aip.org/pacs/pacs2010/individuals/pacs2010_regular_edition/index.html
%%61.43.-j       Disordered solids
%%64.60.Bd       General theory of phase transitions
%%64.70.kj       Glasses
%%64.70.P-       Glass transitions of specific systems     (Heuer,Schroeder)
%%64.70.Q-       Theory and modeling of the glass transition
%45.70.-n 	Granular systems (see also 05.65.+b Self-organized systems)
%64.60.-i 	General studies of phase transition
%  64.70.kj 	Glasses 
%02.70.Ns, %Molecular dynamics and particle methods
%{Glasses}
\maketitle

\section{Introduction}
\label{sec:introduction}
 
If a glass-forming liquid is cooled from a 
high temperature to a low temperature and 
crystallization is avoided, 
the relaxation times of the system increase dramatically.
Depending on the experimental (or simulation) time accessible
in comparison with this growing relaxation time, either
a supercooled liquid (in equilibrium) or a 
glass (out of equilibrium) is
observed~\cite{glassbook,berthierbiroli2011,AngellScience1995}. 
In the non-equilibrium (aging) case, after a temperature quench, 
the dynamics at low temperature 
depends on the waiting time $t_{\rm w}$ -- the time 
elapsed since the temperature quench.
To investigate this rich 
dynamics a large variety of approaches (experiments, computer
simulations and theoretical techniques) have been used
and many different systems have been studied.
Previous work on the dynamics both of supercooled 
liquids and of glasses 
range from small molecules, to polymers, to network glasses,
to colloidal glasses, and to 
granular systems (in the last two cases density is the control
parameter, instead of temperature).
For reviews we refer the reader to
Refs.\cite{hunterweeksreview,berthierbiroli2011,glassbook,leshouches2002,AngellScience1995}.

%\medskip
A common finding of these studies, and in that sense a universal feature,
is that the dynamics is spatially heterogeneous, meaning that there are 
fast and slow regions in space~\cite{berthierdynhetbook,Ediger2000,sillescureview}.
One route for probing the extent of universality is to investigate the 
possible presence of similar scaling behaviors of the dynamical
heterogeneities in diverse systems. We take this route 
in the work presented here. Specifically, we investigate 
the scaling of 
%the global incoherent intermediate scattering function $C(t_{\rm w},t_{\rm w}+t,q,\alpha)$,
the dynamic susceptibility $\chi_4$, 
and of the distribution $P(f_{{\rm s},{\mathbf r}})$ of the 
local incoherent intermediate scattering function $f_{{\rm s},{\mathbf r}}$.
%----
In previous work there have been relatively few studies 
on $P(f_{{\rm s},{\mathbf r}})$ 
\cite{castilloprl88,castilloprb68,chamonJCP121,castilloNatPhys2007,parsaeian09,
Schoepe2013}.
%Hans-Joachim Schoepe (via Palberg-webpage see Schoepe now
%Univ. Thueringen)
% *** Please add (last in list)
% Space-resolved dynamic light scattering probing inhomogeneous
% dynamics in soft matter
% Sebastian Golde, Markus Franke, and Hans Joachim Sch??pe
% AIP Conf. Proc. 1518, 304 (2013); doi: 10.1063/1.4794587
% ****
Also, most previous work on $\chi_4$ 
has focused on the dependence of
$\chi_4$ on the temperature (or density) 
in the supercooled liquid regime
(see  
%Dyn.Het.book pages 80 ff
\S 3.2.4.3 of Ref.\cite{berthierdynhetbook} and for SiO$_2$ specifically see
Refs.\cite{berthierJCP2007I,berthierJCP2007II,berthierpre2007,Vogelpre2004})
There have been fewer 
studies for the aging dynamics, i.e. the dependence of $\chi_4$ on the 
waiting time $t_{\rm w}$, which we discuss in  this
paper~\cite{Parisi1999,parsaeian09,parsaeian08,parsaeianarXiv2008,oukris2010,maggi2012,smessaert2013,gupta2014}.  
% *** Please add (first in list)
% Parisi, G. An increasing correlation length in off-equilibrium
% glasses. J. Phys. Chem. B 103, 4128-4131 (1999)
% ***
%MaggiPRL109,097401
%Gupta PRE90,012137(2014)   chi4/chi4max(1-C)
%\cite{oukris2010}. %chi_4(C)

%\medskip
Predictions for the scaling of $P(f_{{\rm s},{\mathbf
    r}})$ with respect to $t_{\rm w}$ follow from a theoretical
framework for the aging dynamics that explains dynamical heterogeneities
in terms of the presence of Goldstone modes associated with a
broken continuous symmetry under time
reparametrizations~\cite{castilloprl88,chamonprl89,castilloprb68,
  chamonJCP121,chamonJStatMech2007,castilloprb78,mavimbelaJStatMech,
  avilaprl2011,avilapre2013,mavimbelaarXiv2013}.
%--
% In an aging system, two time correlation functions $C(t_{\rm w},t_{\rm
% w}+t)$ break time translation invariance, i.e. $C(t_{\rm w},t_{\rm
% w}+t)$ depends not only on the time difference, $t$, but also on
% $t_{\rm w}$. 
To study the dynamical heterogeneity, i.e. the local fluctuations in the
relaxation, we focus on a local two-time correlation, the local
incoherent intermediate scattering function $f_{{\rm s},{\mathbf
    r}}(t_{\rm w},t_{\rm w}+t)$, which depends on the position
$\mathbf r$, the waiting time $t_{\rm w}$, and the time interval
$t$. The Goldstone modes correspond to space
dependent shifts of the time variable $t \to \phi_{\mathbf r}(t)$ such
that
\begin{equation}
\label{eq:treparametrization}
f_{{\rm s},{\mathbf r}}(t_{\rm w},t_{\rm w}+t) 
    \approx C(\phi_{\mathbf r}(t_{\rm w}),\phi_{\mathbf r}(t_{\rm w}+t)).
\end{equation}
Here $C(t_{\rm w},t_{\rm w}+t)$ corresponds to the global two-time
correlation function \cite{castilloprl88,avilapre2013}. More
generally, a simple Landau-theory approximation for the dynamical
action predicts that quantities describing fluctuations in the system
depend on the waiting time $t_{\rm w}$ and the time interval $t$
essentially only through the global two-time correlation function
$C(t_{\rm w},t_{\rm w}+t)$~\cite{castilloprl88,
  castilloprb68,chamonJCP121}.  Thus it is expected that the probability distribution
$P(f_{{\rm s},{\mathbf r}}(t_{\rm w},t_{\rm w}+t))$ should collapse for
different waiting times $t_{\rm w}$, for $(t_{\rm w},t)$ pairs chosen
such that $C(t_{\rm w},t_{\rm w}+t)$ is held fixed.  This prediction
is consistent with spin glass simulation results 
\cite{chamonJCP121,castilloprb68,castilloprl88}.
%  and experimentally \cite{herissonEPJB2004,herissonPRL2002}.  
Despite the theory being initially derived for spin glasses,
simulation results for structural
glasses~\cite{avilapre2013,avilaprl2011,parsaeian09,parsaeianarXiv2008,parsaeian08,castilloNatPhys2007,gupta2014}
and experimental results for a polymer glass \cite{oukris2010} find
this predicted scaling to hold. The numerical simulations also show
that $\chi_4(t_{\rm w},t_{\rm w}+t)$ is a product of two factors: a
waiting-time dependent scale that grows with $t_{\rm w}$, and a
scaling function that depends on $(t_{\rm w},t)$ only through the
value of $C(t_{\rm w},t_{\rm w}+t)$.

%\medskip
The large variety of structural glass formers can be divided into two
broad groups, called fragile and strong glass formers, due to their
different dependence of the viscosity (and the relaxation time) on
temperature \cite{AngellScience1995,berthierbiroli2011,glassbook}.
All of the previous tests of predictions of the Goldstone mode
approach in structural glasses have been for the case of fragile glass
formers. It is an open question whether the behavior of dynamical
heterogeneity in strong glass formers is also well described by
the same theoretical framework. In this paper we address precisely
that question. We present here molecular dynamics simulation results
for the network former SiO$_2$ which is a strong glass former.  The
van Beest-Kramer-van Santen (BKS) potential \cite{beest_90} which we
use has not only been shown to be an excellent model for real silica
\cite{vollmayr96_2,BKSTc,Badro1997,Taraskin1999} but also previous
work of the last 20 years provides us with detailed insight into many
of the properties
of this system, including its phase
diagram
\cite{lascaris2015,lascaris2014,rajappa2014,farrow2011,poole04,poole01,Badro1998,BarratBadro1997},
energy landscape
\cite{saksaengwijit2004,reinisch2005,saksaengwijit2006,saksaengwijit2007,reinisch2006},
structure
\cite{kvlPRE2013,lascaris2014,rajappa2014,vinh2012,BKSTc,vollmayr96_2}
%defects saksaengwijit2004 specific heat
\cite{lascaris2015,scheidler2001}, vibrational spectrum
\cite{Taraskin1997,Taraskin1999,Taraskin2002,UchinoTaraskin2005,Leonforte2011},
dynamical heterogeneities
\cite{kawasaki2014,kawasaki2013,Vogelpre2004,Vogelprl2004,Bergroth2005,Teboul2006,Hung2013},
and aging \cite{helfferichepl2015,berthier07,kvl2010,kvlPRL2013}
\footnote{This is not  a complete list of BKS-simulations.
For further work please see references therein.}.

%\medskip
Further motivation for the present work is the
unexpected similarity that has recently been found between
the dynamics of the strong glass former SiO$_2$ 
and fragile glass formers \cite{kvlPRL2013}. Whereas in
Ref.~\cite{kvlPRL2013} the  microscopic dynamics is studied
via single particle jump analysis, we investigate in this paper 
whether this surprising similarity of strong and fragile 
glass dynamics also holds true for the scaling of dynamical
heterogeneities. We find not only that indeed most results 
confirm universal dynamics, but we also gain deeper insight 
into the involved length and time scales.
The scaling of $P(f_{{\rm s},{\mathbf r}})$ uncovers the
importance of intermediate length scales and the scaling 
of $C$, $\chi_4$ and $P(f_{{\rm s},{\mathbf r}})$ all indicate
a common aging clock which is the same for Si and O atoms.

%-----------------

\section{Model and Simulation Details}
\label{sec:simulation}

To model amorphous SiO$_2$ we used the BKS potential 
\cite{beest_90}. We carried out molecular dynamics (MD)
simulations with $N_{\rm Si}=112$ silica atoms and $N_{\rm O}=224$
oxygen atoms, at a constant volume 
$V=\left ( 16.920468 \mbox{\AA} \right )^3$
which corresponds to a density $\rho=2.323$ g/cm$^3$.
For further details on the interaction see Ref.\cite{kvl2010}.

%see ~/student_research.dir/chris.dir/sourcefiles.dir/runprep5000_stoch.pl
%see ~/student_research.dir/chris.dir/sourcefiles.dir/parfile_6000K_run1
At $6000$~K we generated 200 independent configurations
(at least $1.63$~ns apart) which
%(at least $10^6$MDsteps*0.16*1.0217D-5ns), which 
then were fully equilibrated at initial temperature 
$T_{\rm i} = 5000$~K 
for $3.27$~ns, followed by 
%(for ($2\cdot 10^6$MDsteps)*0.16*1.0217D-5), followed by 
an instantaneous quench to lower temperature 
$T_{\rm f} = 2500$~K,
i.e.  below $T_c=3330$~K.
Unique to our simulations is that we applied the
Nos\'e-Hoover temperature bath at $T_{\rm f}$ only 
for the first $0.327$~ns (NVT) 
%(same as in PRL 320000*0.1*1.0217D-5ns)
and then continued with constant energy (NVE) for $98.1$~ns 
%(60000000*0.16*1.0217D-5ns which is 3times longer than PRL)
to disturb the dynamics minimally.  
We confirmed that $T_{\rm f}$ stays constant and is similar 
to $T_{\rm f}(t)$ as shown in Fig.~2 of Ref.\cite{kvl2010}.
The MD time step was $1.02$~fs and $1.6$~fs during the (NVT)
and (NVE) runs respectively.
In what follows we analyzed
the combined (NVT) and (NVE) simulation runs at $T_{\rm f}$.

The main difference between the present simulation and the ones
discussed in Refs.~\cite{kvl2010,kvlPRL2013,kvlPRE2013} is that our new
dataset has increased statistics (200 independent runs instead of 20)
and that each NVE run at $T_{\rm f}$ has a longer duration (98.1 ns
instead of 32.7 ns).
% and the simulation runs presented here (for $T_{\rm f}=2500$~K) is
% that we ran 200 instead of 20 independent simulation runs and that
% the (NVE) run at $T_{\rm f}$ was for $98.1$~ns (instead of
% $32.7$~ns). 
As described in section Sec.~\ref{sec:Fs}, this increased statistics
and the longer simulation runs allowed us to gain insight into the
scaling of the two point correlation function $C^{\alpha}$
significantly beyond the results of Ref.~\cite{kvl2010}.
Furthermore, having 200 independent simulation runs gave good enough
statistics to make it possible to determine the dynamic susceptibility
and the distribution of the local incoherent intermediate scattering
function (see Secs~\ref{sec:chi4} and \ref{sec:pofcq}).

\section{Global Incoherent Intermediate Scattering Function}
\label{sec:Fs}

Let us first look at the global generalized incoherent intermediate
scattering function
\begin{equation}
C^{\alpha}(t_{\rm w},t_{\rm w}+t,q) = \left \langle 
f_{\rm s}^{\alpha}(t_{\rm w},t_{\rm w}+t,{\mathbf q}) 
\right \rangle, 
\label{eq:Cq}
\end{equation}
with 
\begin{equation}
f_{\rm s}^{\alpha}(t_{\rm w},t_{\rm w}+t,{\mathbf q})=
    \frac{1}{N_{\alpha}} \sum \limits_{j=1}^{N_{\alpha}}
      \cos \left \{ {\mathbf q} \cdot
  \big ({\mathbf r}_j(t_{\rm w}+t)-{\mathbf r}_j(t_{\rm w})\big
  )\right \}.
\label{eq:fs}
\end{equation}
Here ${\mathbf r}_j(t)$ is the position of particle $j$ at time $t$,
$t_{\rm w}$ is the waiting time elapsed since the temperature quench
from $5000$~K to $2500$~K, and $N_{\alpha}$ is the total number of
particles of type $\alpha$ ($\alpha \in \{{\rm
  Si,O,\textrm{all}}\}$). The notation $\left\langle \ldots
\right\rangle$ indicates an average over wave vectors ${\mathbf q}$ of
fixed magnitude $q$ and over the 200 independent simulation runs.  In
Eq.~(\ref{eq:fs}) the sum is over particles in the complete simulation
box, whereas in Sec.~\ref{sec:pofcq} the sum is only over particles
within a local sub-box. We call $C^{\alpha}$ the ``global'' incoherent
intermediate scattering function to stress this distinction. Error
bars for $C^{\alpha}$ are given by the statistical error of the
average over the 200 independent simulation runs. Even though
averaging over 200 independent runs allows us to remove a lot of the
noise in the results, we further smooth the results by additionally
applying a time average. The time average is computed by using
logarithmic time bins and by averaging $t$-values, $C^{\alpha}$-values
and error bars $\Delta C^{\alpha}$ within the same time bin.

%*** average t-values??? ***

Unique to the present work is that we investigate directly
the influence of particle type on scaling. To do so, we distinguish
three different cases, labeled by the symbol $\alpha$. In the case of
$\alpha=$Si the sum in Eq.~(\ref{eq:Cq}) is exclusively over Si atoms
and in the case of $\alpha=$O the sum is exclusively over O atoms. In
the third case, $\alpha=$all, the sum is over all particles,
i.e. including both Si and O atoms. We use this notation throughout
the whole paper, including in our discussion of the dynamic
susceptibility $\chi_4$ and the probability distribution $P(f_{{\rm
    s},{\mathbf r}})$.
%----

%-------------
\begin{figure}[h]
\includegraphics[width=3.4in]{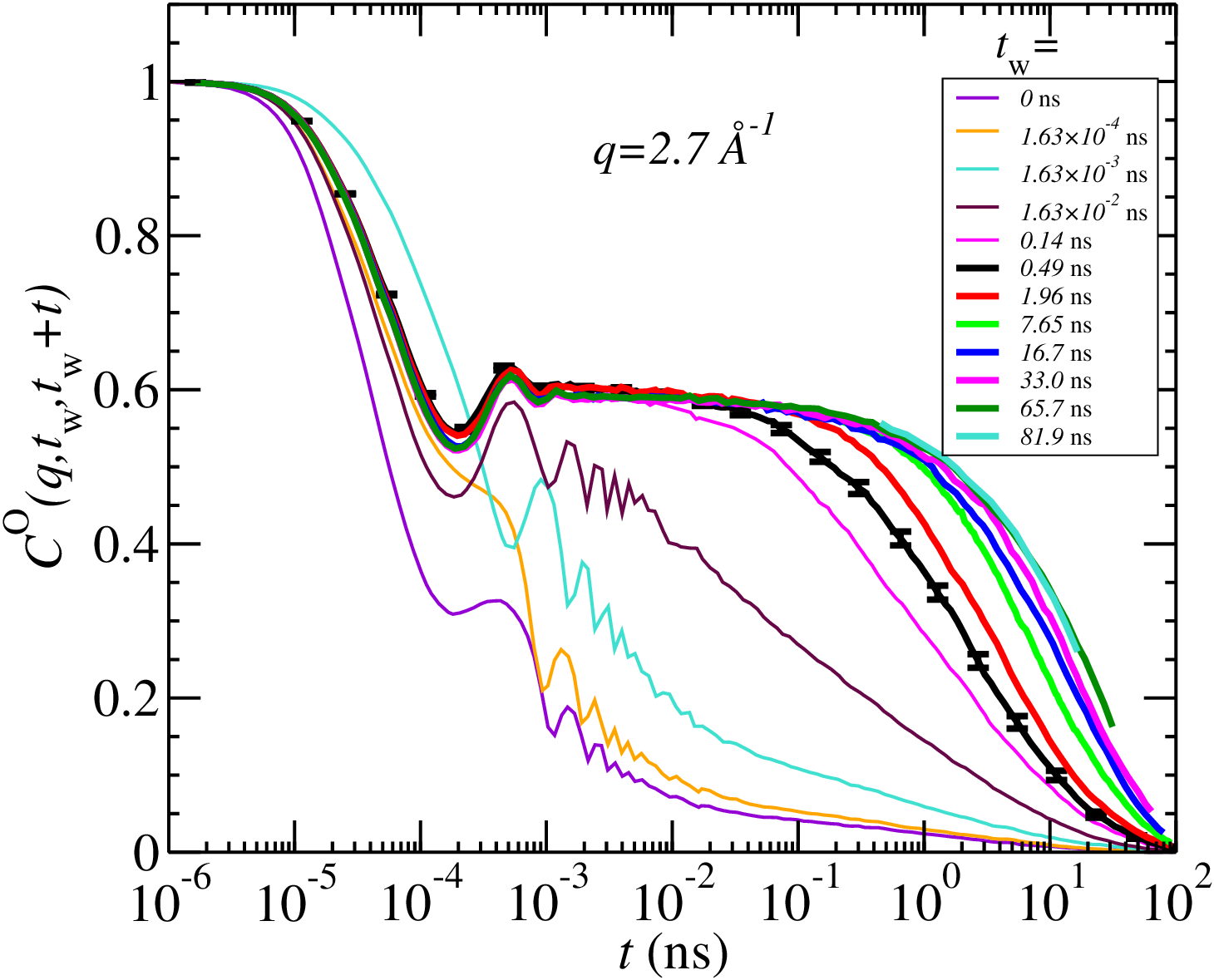}
\caption{\sf (color online) Global generalized incoherent intermediate
  scattering function at $q=2.7 \AA^{-1}$ for oxygen atoms, $C^{\rm
    O}(t_{\rm w},t_{\rm w}+t,q)$ as defined in Eq.~(\ref{eq:Cq}), for
  12 waiting times $t_{\rm w}$ between $0$ ns and $81.9$ ns. To avoid
  cluttering the graph, in this figure and in most other figures where
  results for different waiting times are compared, statistical error
  bars are shown for just one of the waiting times.}
\label{fig:fslntav_q27_O}
\end{figure}
%-------------

Fig.~\ref{fig:fslntav_q27_O} shows $C^{\alpha }(t_{\rm w},t_{\rm w}+t,q)$ 
for oxygen atoms and for $q=2.7 \AA^{-1}$. We find that with increasing 
waiting time $t_{\rm w}$ the correlation function decays more slowly.
To quantify this we
define the relaxation time $\tau^{\alpha}_q$ as the time when $C_{\alpha}$ 
has decayed to a certain value $C_{\rm cut}^{\alpha}(q)$ 
\begin{equation}
C^{\alpha}(t_{\rm w},t_{\rm w}+\tau^{\alpha}_q,q)=C_{\rm cut}^{\alpha}(q)
\hspace*{2mm}{\mbox .}
\label{eq:trCq}
\end{equation}

Instead of the commonly used choice $C_{\rm cut}^{\alpha}=1/e$, we
adjust to the varying plateau height of $C^{\alpha}(q)$ for different
$q$.  Hence for $C_{\rm cut}^{\alpha}(q)$ we choose the values listed
in Table~\ref{table:Cqcut}, each of which is given by $1/e$ times the
corresponding plateau value of $C^{\alpha}(q)$.

\begin{table}[h]
\centering% NICHT \begin{center}
\begin{tabular}{|c|c|c|c|} \hline \hline
  $q$ & Si & O & \textrm{all}  \\ \hline \hline
1.7 & 0.323 & 0.298  & 0.306  \\ \hline
2.7 & 0.265  & 0.217  & 0.233  \\ \hline
3.4 & 0.221  & 0.162  & 0.182 \\ \hline
4.6 & 0.144 & 0.085 & 0.105 \\ \hline
\end{tabular}
\caption{$C_{\rm cut}^{\alpha}(q)$ values}
\label{table:Cqcut}
\end{table}

%-------------
\begin{figure}[h]
\includegraphics[width=3.4in]{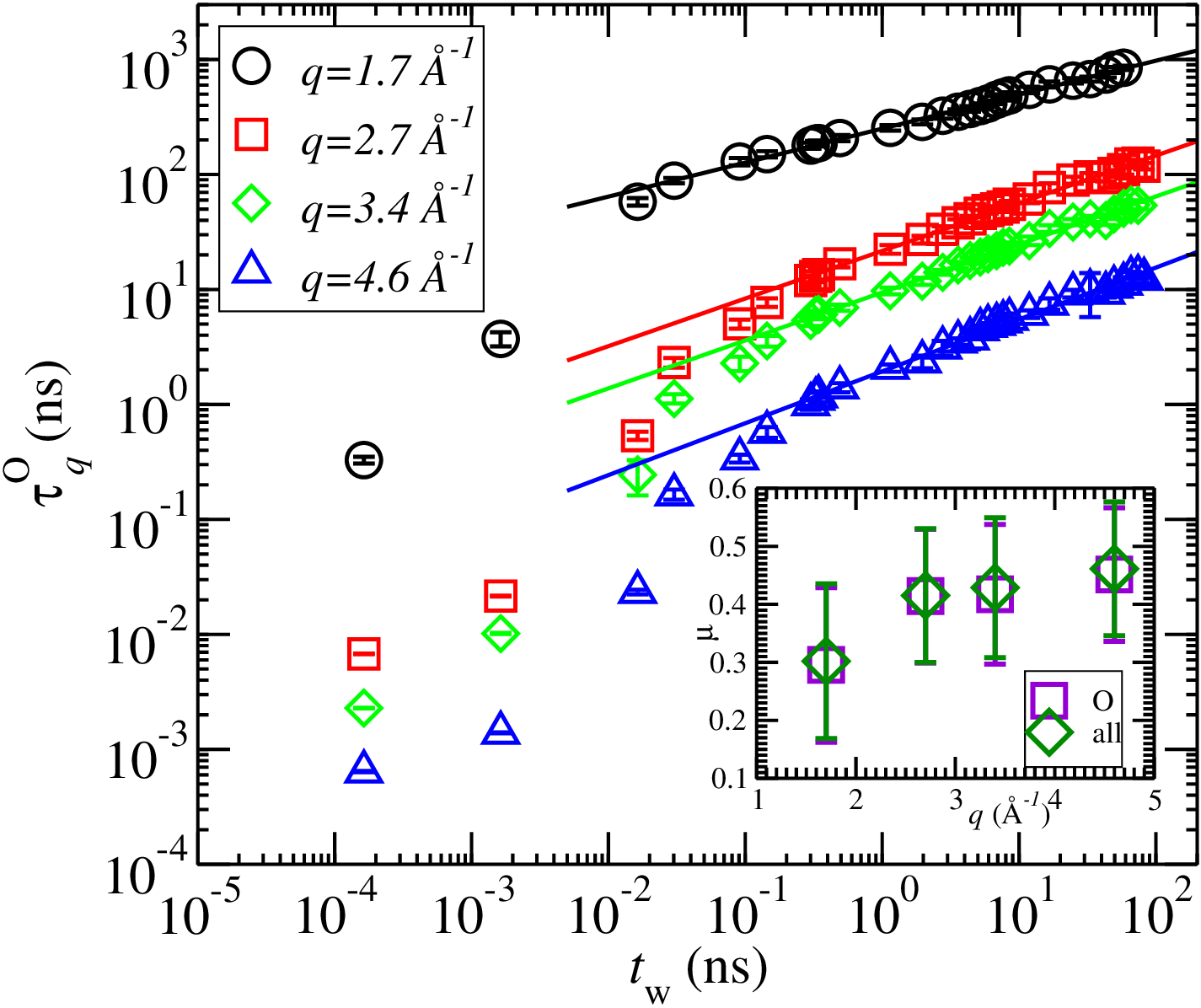}
\caption{\sf (color online) Relaxation times for oxygen atoms
  $\tau^{\rm O}_q(t_{\rm w})$ as defined in Eq.~(\ref{eq:trCq}). For
  clarity the data for $q=4.6 \AA^{-1}, 3.4 \AA^{-1}, 2.7 \AA^{-1}$
  and $1.7 \AA^{-1}$ have been shifted by factors of $1,3,9,27$
  respectively. The lines are power law fits.
      %$\tau^{\alpha}_q(t_{\rm w})={\rm e}^A t_{\rm w}^{\mu}$,   
      The fitted exponents $\mu$ are shown in the inset, both for the
      case of O atoms and for the case of all atoms.
}
\label{fig:trCq_O}
\end{figure}
%-------------

The resulting relaxation times for oxygen atoms are shown in
Fig.~\ref{fig:trCq_O} as functions of the waiting time $t_{\rm w}$. As
in Ref.~\cite{kvl2010}, different regimes for $t_{\rm w}$ can be
identified from Figs.~\ref{fig:fslntav_q27_O} and~\ref{fig:trCq_O}.
In Ref.~\cite{kvl2010} it was found that for small waiting times, $t_{\rm
  w} \lesssim 0.1$\ ns, $C^{\alpha}(t_{\rm w},t_{\rm w}+t)$ does not
form a plateau, for intermediate $t_{\rm w}$ a plateau is formed and
time superposition applies, and for sufficiently large waiting times
$C^{\alpha}$ becomes $t_{\rm w}$-independent, i.e. equilibrium is
reached.  The increased statistics of the present simulations allow
the identification in Fig.~\ref{fig:trCq_O} of the transition from
small to intermediate $t_{\rm w}$ as a change in the $t_{\rm w}$
dependence of the relaxation times $\tau^{\alpha}(t_{\rm w})$ from
non-power law to power law behavior.  Fig.~\ref{fig:trCq_O} also shows
that $\tau^{\alpha}_q(t_{\rm w})$ does not reach a plateau, i.e. the
waiting times are not long enough to reach equilibrium.

%\medskip
Let us next investigate further the dynamics for intermediate waiting
times $t_{\rm w} \gtrsim 0.1$\ ns.  The inset in Fig.~\ref{fig:trCq_O}
shows the power law fit exponents $\mu$ as functions of wave vector
$q$. We find that within the error bars, $\mu$ seems to be independent
of $q$ for $q \ge 2.7 \AA^{-1}$.  Similar results for $\tau(t_{\rm
  w})$ have been found experimentally for a metallic glass
\cite{ruta2013} and for a colloidal glass \cite{marques2015}.  The
inset in Fig.~\ref{fig:trCq_O} also shows that $\mu$ is independent of
the particle type $\alpha$. (The particle type is indicated by a
square for $\alpha=$O and by a rhombus for $\alpha=$all.)  This
independence of $\alpha$ is rather surprising, since Horbach and Kob
had found that the dynamics of silicon and oxygen atoms is very
different for temperatures below $3330$~K \cite{BKSTc}. Saksaengwijit
and Heuer \cite{saksaengwijitPRE74_2006} relate this decoupling of
silicon and oxygen dynamics to rotational processes.
% **** Is this because those previous works focused on other time
% regimes??? *** 

%\medskip
We interpret the $\alpha$-independence of $\mu$ as an evidence for the
existence of a common ``single aging clock'' in the system, despite
the different dynamics of Si and O atoms. This allows us to analyze Si
and O atoms together ($\alpha=$all). To {\em directly} test the
hypothesis of a single aging clock, we generalize an approach
introduced by Kob and Barrat \cite{KobBarratCq1Cq2}. They had
investigated the $q$-dependence of $C^{\alpha}$ via a parametric plot
of $C^{\alpha}(q_2)$ versus $C^{\alpha}(q_1)$ for various $t_{\rm w}$.
Whereas for their system, a binary Lennard-Jones system, they found no
data collapse \cite{KobBarratCq1Cq2}, for our system, SiO$_2$, it was
found in Ref.~\cite{kvl2010} that data collapse indeed happens. This
indicates that $C^{\alpha}(t_{\rm w},t_{\rm w}+t,q) =
C^{\alpha}(\tilde{z}(t_{\rm w},t,\alpha),q)$~\cite{kvl2010}. In
other words, for each particle type $\alpha$ there is a unique
$q$-independent aging clock represented by $\tilde{z}(t_{\rm
  w},t,\alpha)$. Notably, this is the only non-universal result we encounter
in our comparison of the dynamics of fragile and strong glass
formers. We now address the question of the existence of a common
aging clock for Si and O atoms, in other words, whether or not the
function $\tilde{z}(t_{\rm w},t,\alpha)$ is independent of
$\alpha$. The question is therefore whether it is true that
\begin{eqnarray}
\label{eq:Cqofqalpha}
C^{\alpha}(t_{\rm w},t_{\rm w}+t,q)& \equiv &
C(t_{\rm w},t_{\rm w}+t,q,\alpha) \nonumber \\ %\nonumber
&{? \atop =}& C(z(t_{\rm w},t),q,\alpha)
\hspace*{2mm}\mbox{.}
\end{eqnarray}

To answer this question we investigate directly the
$\alpha$-dependence via a parametric plot of $C^{\rm O}(t_{\rm
  w},t_{\rm w}+t,q)$ versus $C^{\rm Si}(t_{\rm w},t_{\rm w}+t,q)$ as
shown in Fig.~\ref{fig:fsOoffsSi_q27q17} for $q=2.7 \AA^{-1}$ (and in
the inset for $q=1.7 \AA^{-1}$). We find almost perfect data collapse
for all investigated $q$, and conclude that Eq.~(\ref{eq:Cqofqalpha})
is correct. Hence the $(t_{\rm w},t)$-dependence is solely governed by
one function $z(t_{\rm w},t)$, i.e. an ``inner aging clock'' which is
not only $q$-independent but also the same for different particle
types.

%-------------
\begin{figure}
\includegraphics[width=3.4in]{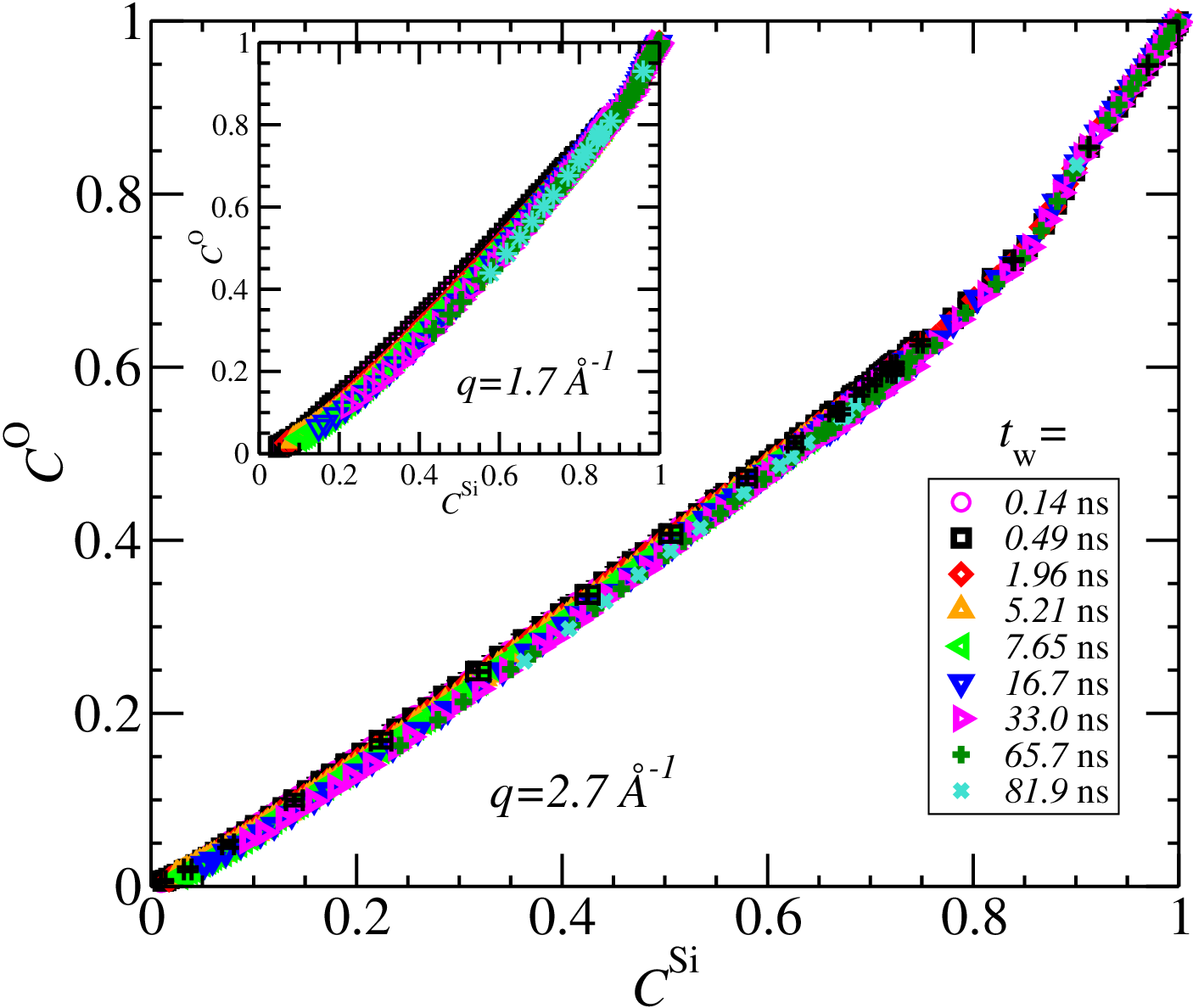}
\caption{\sf 
(color online) 
To study directly the dependence of the global incoherent 
intermediate scattering function $C^{\alpha}$ on the particle 
type $\alpha$, we show here two parametric plots of
$C^{\rm O}$ versus $C^{\rm Si}$ for various waiting times $t_{\rm w}$:
one for $q=2.7 \AA^{-1}$ (main panel) and another for $q=1.7 \AA^{-1}$
(inset).
}
\label{fig:fsOoffsSi_q27q17}
\end{figure}
%-------------

This may appear surprising at first, since, as mentioned before, it is
known that in SiO$_2$ the oxygen atoms have a faster dynamics than the
silicon atoms~\cite{BKSTc, saksaengwijitPRE74_2006}. Our results do
not contradict this statement. 
To illustrate how to reconcile a common clock and yet different
dynamics of Si and O atoms, we show in Fig.~\ref{fig:clock}
$(t_{\rm w},t_{\rm w}+t)$-pairs for fixed $C^{\alpha}$. 
For example, the blue circles of the bottom curve were obtained 
by finding for each $t_{\rm w}$ the corresponding 
$t_{\rm w}+t$ for which 
$C^{\rm Si}(t_{\rm w},t_{\rm w}+t,q=2.7 \AA^{-1})=0.575$
with $1$\% accuracy. 
For equilibrium dynamics, this curve would be trivial: it would 
be the set of $(t_{\rm w},t_{\rm w}+t)$-pairs  for a certain constant 
value of $t$. For aging dynamics, the curve
is non-trivial: $t$ changes as $t_{\rm w}$ changes.
Nevertheless, we obtain the same non-trivial curve for 
O atoms  (blue triangles) for fixed 
$C^{\rm O}=0.463$. Similarly we obtain identical curves for Si atoms 
(circles) and oxygen atoms (triangles) for different choices of 
$C^{\rm Si}$ and $C^{\rm O}$ (upper three curves), hence a common 
clock for Si and O atoms.
The fact that the dynamics of O atoms is faster than the
dynamics of Si atoms is reflected by the constant value of $C^{\rm O}$
being lower than the corresponding constant value of $C^{\rm Si}$ 
on the same curve.

%-------------
\begin{figure}
\includegraphics[width=3.4in]{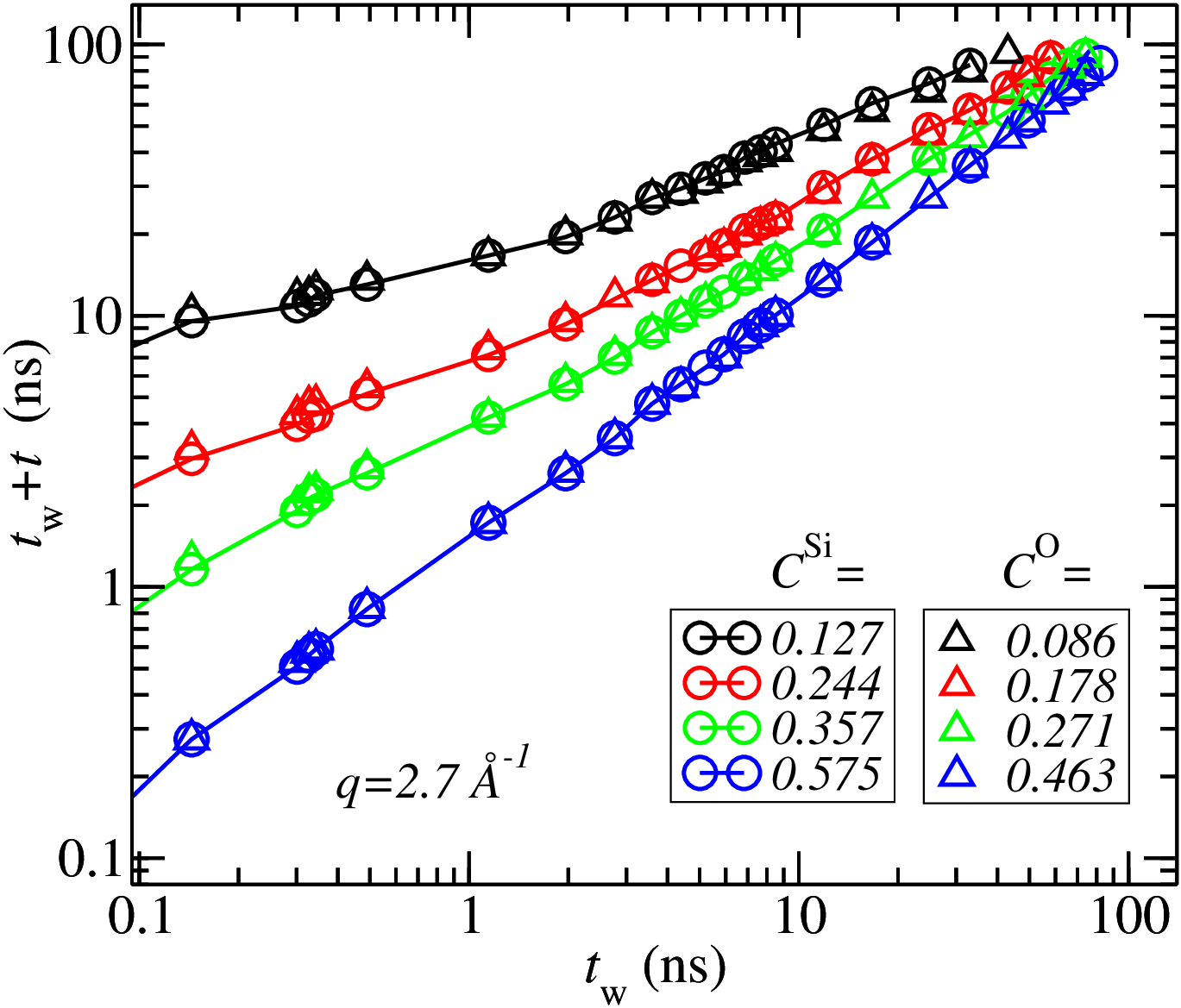}
\caption{\sf 
(color online) 
This figure illustrates the compatibility of Si and O atoms 
having both different speeds as well as a common clock.
For each $t_{\rm w}$ the corresponding $(t_{\rm w}+t)$ was 
determined to obtain the specified 
$C^{\alpha}=C^{\alpha}(t_{\rm w},t_{\rm w}+t,q=2.7 \AA^{-1})$
with $1$\% accuracy. The solid lines are for the guidance of the eye.
The common clock is apparent by identical curves for silicon (circles) 
and oxygen (triangles).
The fact that the dynamics of O atoms is faster than the
dynamics of Si atoms is reflected by the constant value of $C^{\rm O}$
being lower than the corresponding constant value of $C^{\rm Si}$ 
on the same curve.
}
\label{fig:clock}
\end{figure}
%-------------

\section{Dynamic Susceptibility}
\label{sec:chi4}

In this section we study the dynamic susceptibility,
$\chi_4^{\alpha}$, which is a four-point correlation function.
$\chi_4^{\alpha}$ quantifies thermal fluctuations of the incoherent
intermediate scattering function.  Following the notation of Berthier
\cite{berthierpre2007} $\chi_4^{\alpha}$ is defined~\footnote{ In the
  ensemble used in our simulations the numbers of both O and Si atoms
  are kept constant, and our definition of dynamic susceptibility
  describes fluctuations in that particular ensemble. The dynamic
  susceptibility in a different ensemble could have a different
  value~\cite{berthierJCP2007I}. For example, a possible alternative
  would be an ensemble that allows particle number fluctuations, thus
  yielding an extra contribution that would increase the value of the
  dynamic susceptibility. We do not pursue such alternative in the
  present work.} to be

\begin{equation}
\chi_4^{\alpha}(t_{\rm w},t_{\rm w}+t,q)=N_{\alpha}  \left [
   \left \langle  \left(f_{\rm s}^{\alpha} \right)^2 \right \rangle 
        - \left ( \langle  f_{\rm s}^{\alpha} \rangle \right )^2 \right ]
   \hspace*{0mm}\mbox{,}
\label{eq:chi4oft}
\end{equation}

% \begin{equation}
% \chi_4^{\alpha}(t_{\rm w},t_{\rm w}+t,q)=N_{\alpha}  \left [
%   \left \langle  \left(
%   f_{\rm s}^{\alpha}(t_{\rm w},t_{\rm w}+t,{\mathbf q}) 
%   \right)^2 \right \rangle 
%   - \left ( \langle  
%   f_{\rm s}^{\alpha}(t_{\rm w},t_{\rm w}+t,{\mathbf q}) 
%   \rangle \right )^2 \right ]
%\label{eq:chi4oft}
% \end{equation}

\noindent
where $f_{\rm s}^{\alpha}(t_{\rm w},t_{\rm w}+t,{\mathbf q})$,
$N_{\alpha}$ and $\left \langle \ldots \right \rangle $ are as defined
at the beginning of Sec.~\ref{sec:Fs}. To obtain error bars for
$\chi_4^{\alpha}$ we divide the 200 independent simulation runs into
20 subsets each of 10 independent simulation runs, and compute a value
$\chi_4^{(\alpha, i)}$ for each subset $i$, with $i=1,\cdots,20$. The
error bar of $\chi_4^{\alpha}$ is the standard deviation of the mean
over those 20 independent $\chi_4^{(\alpha, i)}$ values. As in the
case of $C^{\alpha}$, we further smooth the results for
$\chi_4^{\alpha}$ by applying a time average using logarithmic time
bins.  This means that for each logarithmic bin all unsmoothed
data $(t,\chi_4,\Delta \chi_4)$ which occur at a time $t$ within the
specified bin time window, are averaged  to 
$(\overline{t},\overline{\chi_4},\overline{\Delta \chi_4})$.

\begin{figure}
\includegraphics[width=3.4in]{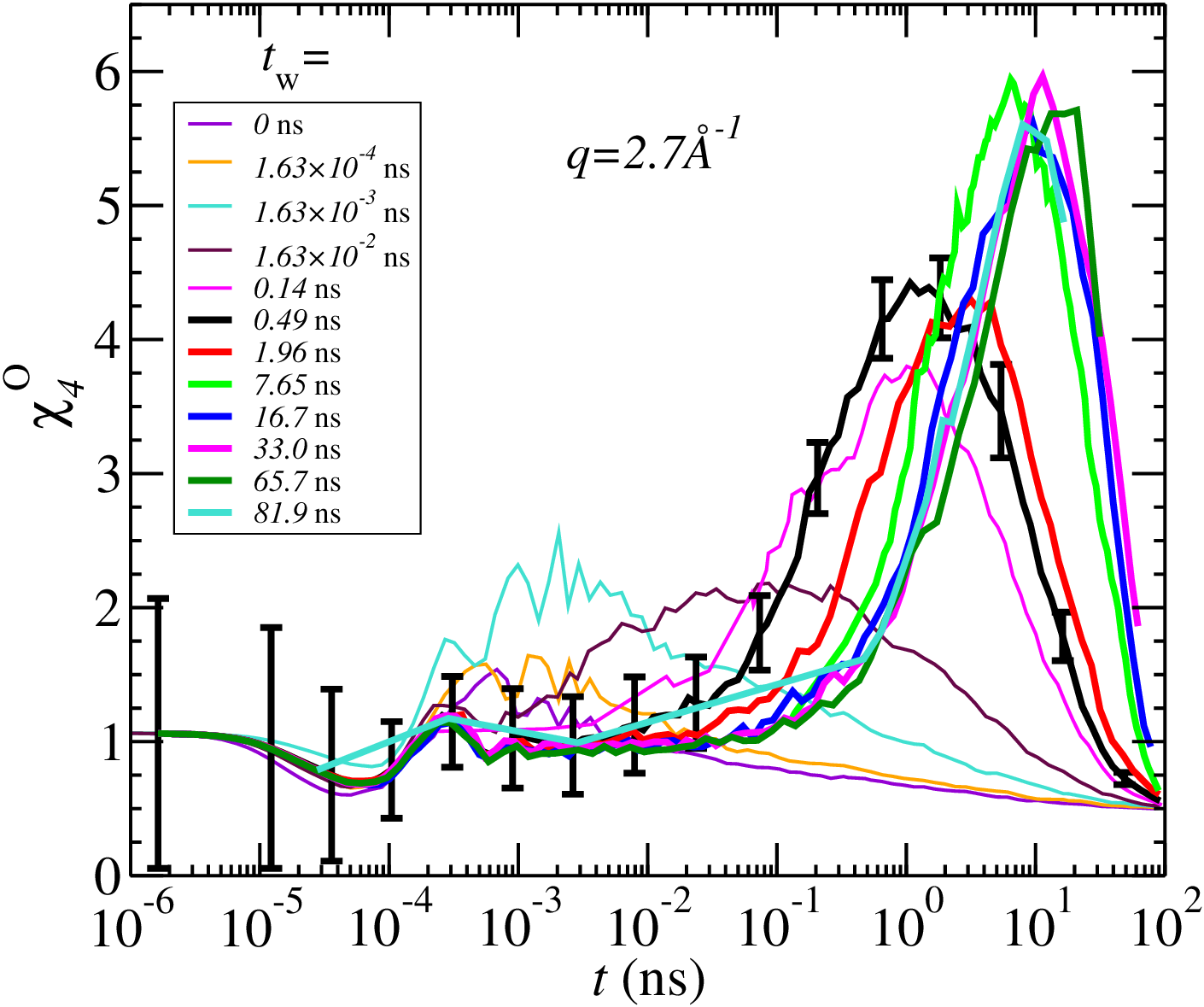}
\caption{\sf (color online) 
      Dynamic susceptibility 
        $\chi_4^{\rm O}(t_{\rm w},t_{\rm w}+t,q)$
      as defined in Eq.~\ref{eq:chi4oft} for $q=2.7 \AA^{-1}$, for the
      same waiting times $t_{\rm w}$ as in
      Fig.~\ref{fig:fslntav_q27_O}. 
}
\label{fig:chi4lntav_O}
\end{figure}
%-------------

%\medskip
Fig.~\ref{fig:chi4lntav_O} shows the resulting dynamic susceptibility
for oxygen atoms and $q=2.7 \AA^{-1}$.  Since $\chi_4^{\alpha}$ is a
four-point correlation function, it is a measure of dynamic
heterogeneities.  The dynamic susceptibility is small both for very
short times and for very large times $t$ and it has a maximum
$\chi_{\rm max}^{\alpha}$ at an intermediate time $t_{\rm
  max}^{\alpha}$.  This maximum can be interpreted as a maximal number
of particles in a dynamically correlated region. For a thorough
discussion of this maximum and its scaling dependence on temperature
and system size in the case of a supercooled liquid and a dense
granular system see
Refs.~\cite{berthierpre2007,berthierJCP2007II,karinaprl2014}.  We
investigate here instead the aging dynamics and thus the dependence of
$\chi_4^{\alpha}$ on waiting time $t_{\rm w}$.  To quantify the
dependence of this maximum on $t_{\rm w}$, $q$ and $\alpha$, we show
the peak height $\chi_{\rm max}^{\alpha}$ and the peak position
$t_{\rm max}^{\alpha}$ as functions of $t_{\rm w}$ in
Figs.~\ref{fig:chi4max_O} and \ref{fig:tmax_O} respectively.  Similar
to previous results in fragile glass
formers~\cite{smessaert2013,parsaeian09,parsaeian08,parsaeianarXiv2008,maggi2012,gupta2014}
$\chi_{\rm max}^{\alpha}$ and $t_{\rm max}^{\alpha}$ increase with
increasing $t_{\rm w}$.
 %$t_{\rm w} \gtrsim 0.1$~ns.  
%
It is possible that $\chi_{\rm max}^{\alpha}$ reaches a plateau for
large $t_{\rm w}$, but the noise in the results is too large to allow
for any definite conclusions to be drawn.  As above, the dependence of
$t_{\rm
  max}^{\alpha}(t_{\rm w},q)$ on $t_{\rm w}$ is consistent with the
presence of two 
regimes: $t_{\rm w} \lesssim 0.1$~ns and $t_{\rm w} \gtrsim
0.1$~ns. As in the case of $\tau^{\alpha}_q(t_{\rm w})$, in the longer
time regime the timescale $t_{\rm max}^{\alpha}(t_{\rm w},q)$ has a
power law dependence on $t_{\rm w}$, with exponents $\mu_{\rm max}$
which are again independent of particle type, as shown in the inset of
Fig.~\ref{fig:tmax_O}.

% KVLnote: I checked in fsqt_cal_30.pl   that max-t,chi4 are 
%    indeed position and height of curves shown in Fig.~\ref{fig:chi4lntav_O}}
%-------------
\begin{figure}[h]
\includegraphics[width=3.4in]{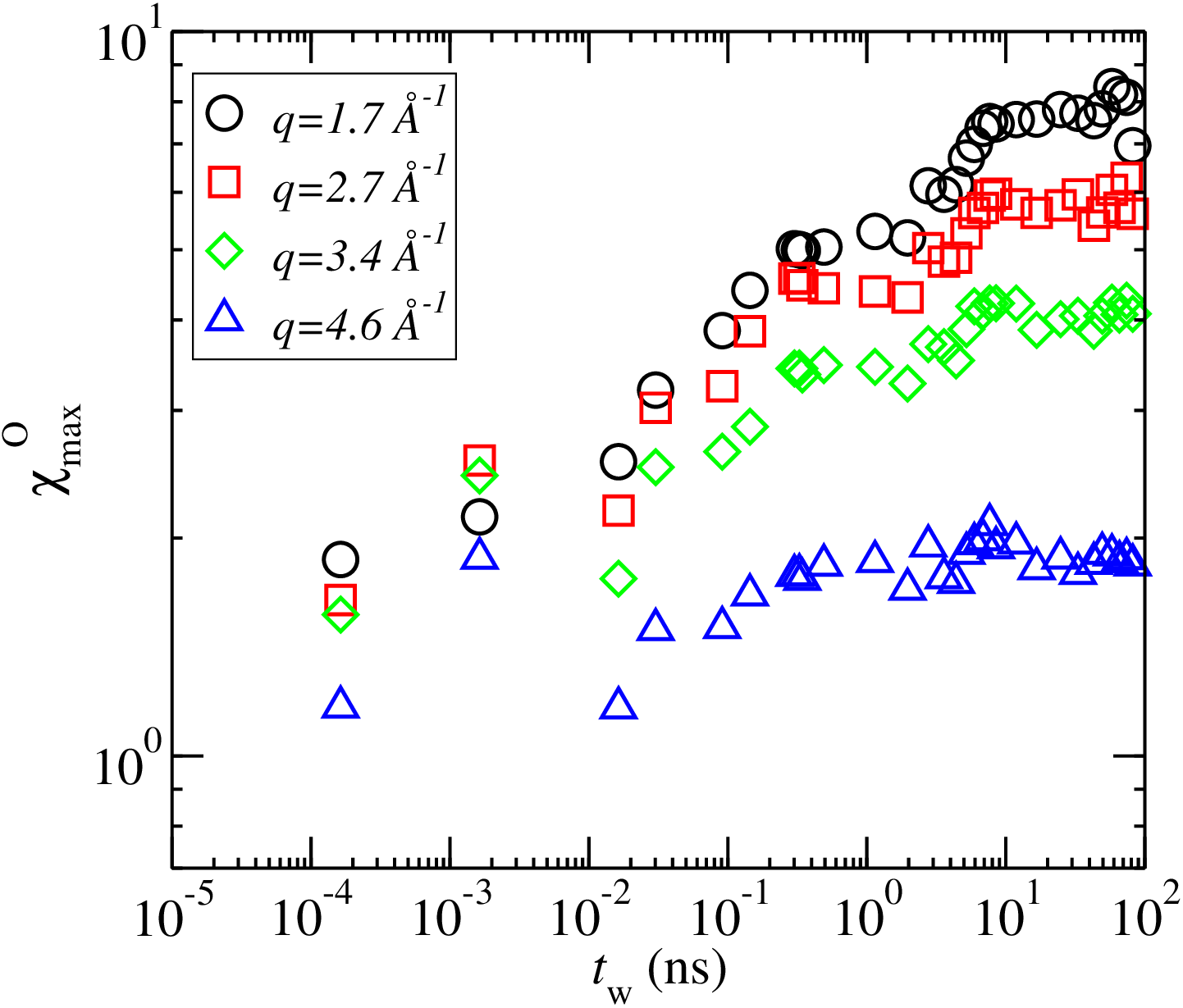}
\caption{\sf (color online) 
     Peak value, $\chi_{\rm max}^{\alpha}$, 
     of 
     % the maximum of 
       $\chi_4^{\alpha}(t_{\rm w},t_{\rm w}+t,q)$ 
     (see Fig.~\ref{fig:chi4lntav_O}).
     Here $\chi_{\rm max}^{\alpha}$ as function of waiting time $t_{\rm w}$
     is shown for oxygen atoms, i.e. $\alpha=$O. 
     We find similar results for $\alpha=$Si
     and $\alpha=$all.
}
\label{fig:chi4max_O}
\end{figure}
%-------------

%-------------
\begin{figure}[h]
\includegraphics[width=3.4in]{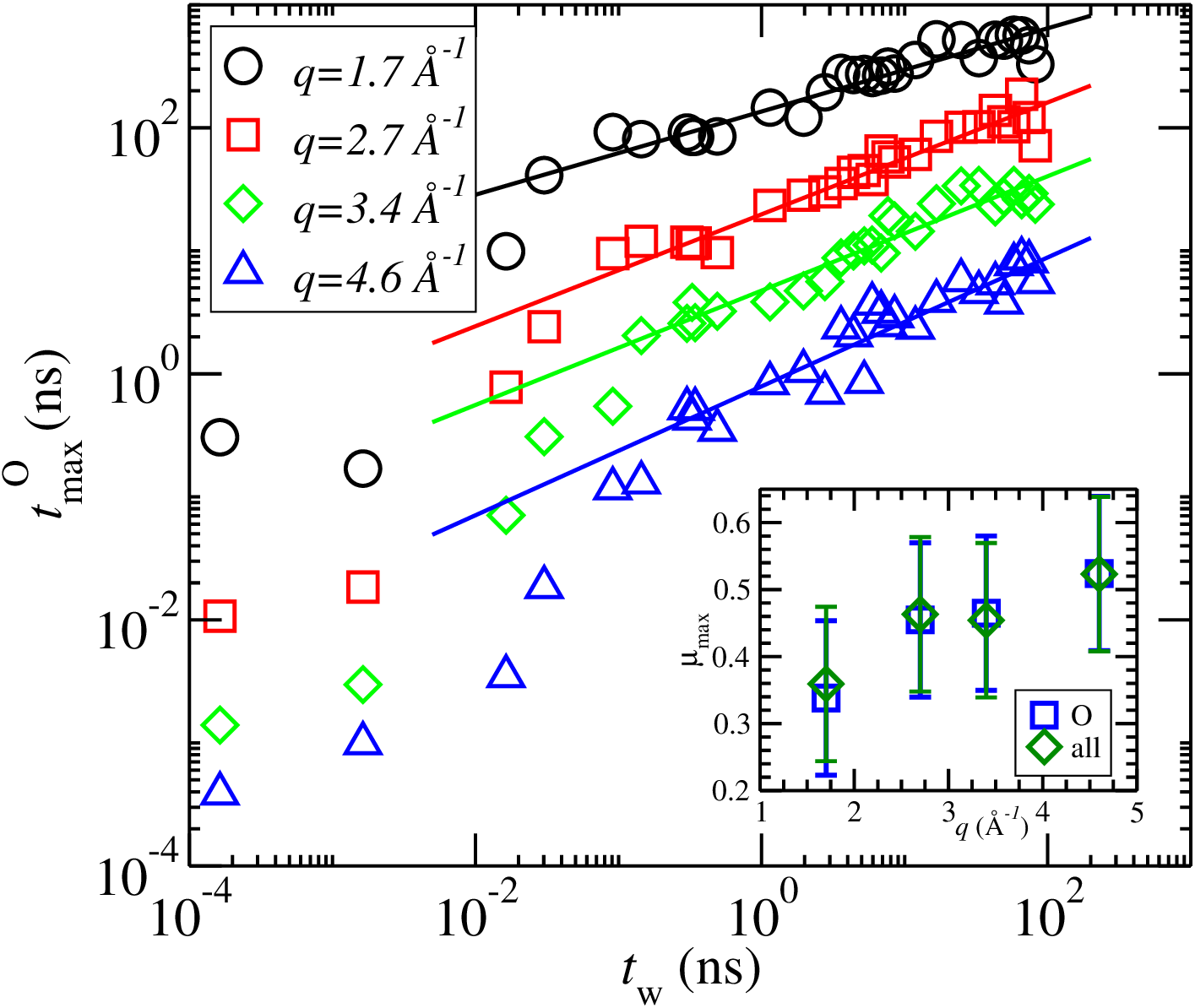}
\caption{\sf (color online) 
     Peak position $t_{\rm max}^{\rm O}$ of the maximum of 
       $\chi_4^{\rm O}(t_{\rm w},t_{\rm w}+t,q)$ (see Fig.~\ref{fig:chi4lntav_O}).
      For clarity the data for $q=4.6 \AA^{-1}, 3.4 \AA^{-1}, 2.7
      \AA^{-1}$  and $1.7 \AA^{-1}$ have been shifted by factors of
      $1,3,9,27$  respectively. 
      % For clarity the data for each of the values have been shifted
      % by factors of $27$, $9$, $3$, $1$. 
     The lines are power law fits $t_{\rm max}^{\rm O} \propto t_{\rm
       w}^{\mu_{\rm max}}$ with exponents as shown in the inset.
    %{\bf Note to KVL: Horacio gave some argument that if C prop exp(-t/const) 
    % then trCq prop tmax  so mu(mumax) along diagonal. I checked this
    %with  paste fitvaluefiles and gawk and it is not valid.}
}
\label{fig:tmax_O}
\end{figure}
%-------------

%\medskip
We next address the question of how the dynamic susceptibility scales
with respect to waiting time. As described in the introduction,
numerical simulations for fragile glasses
find~\cite{parsaeian08,parsaeian09,parsaeianarXiv2008,gupta2014} a
scaling behavior of the $(t_{\rm w},t)$ dependence of the dynamic
susceptibility given by

\begin{equation}
\label{eq:chi4horacio}
\chi_4^{\alpha}(t_{\rm w},t_{\rm w}+t,q)=\chi_4^0(t_{\rm w},q,\alpha) \,
%\phi^{\alpha}(C^{\alpha}(t_{\rm w},t_{\rm w}+t,q),q)
\phi(C^{\alpha}(t_{\rm w},t_{\rm w}+t,q),q,\alpha)
\mbox{.}
\end{equation}

% Our simulations are for SiO$_2$,  we thus test whether
% Eq.(\ref{eq:chi4horacio}) holds also for the strong glass former
% SiO$_2$. 
A similar more general result follows from Ref.~\cite{dyreJCP2015}.
Without loss of generality, we choose $\chi_4^0(t_{\rm w},q,\alpha)$
in Eq.~(\ref{eq:chi4horacio}) to be the maximum height $\chi_{\rm
  max}^{\alpha}$.
To test the validity of Eq.~(\ref{eq:chi4horacio}), we plot in
Fig.~\ref{fig:chi4fsscaling_q27_O} $\chi_4^{\alpha}/\chi_{\rm
  max}^{\alpha}$ as a function of $\left ( 1-C^{\alpha} \right )$ for
the case of oxygen atoms ($\alpha=$O) and $q=2.7
\AA^{-1}$ (for details see endnote \footnote{To preserve scaling, the  
   details of smoothing data needs care. We used the smoothed data of
   Fig.~\ref{fig:chi4lntav_O} to determine $\chi_{\rm max}^{\alpha}$ which
   is identical to the values of Fig.~\ref{fig:chi4max_O}. We then 
   determined for each $t_{\rm w}$ the unsmoothed data 
   $\chi_4^{\alpha}/\chi_{\rm max}^{\alpha}$ as function of
   $(1-C^{\alpha})$. The thus obtained unsmoothed data were then smoothed
   via linear binning of $(1-C^{\alpha})$.}.)
%-------------
\begin{figure}
\includegraphics[width=3.4in]{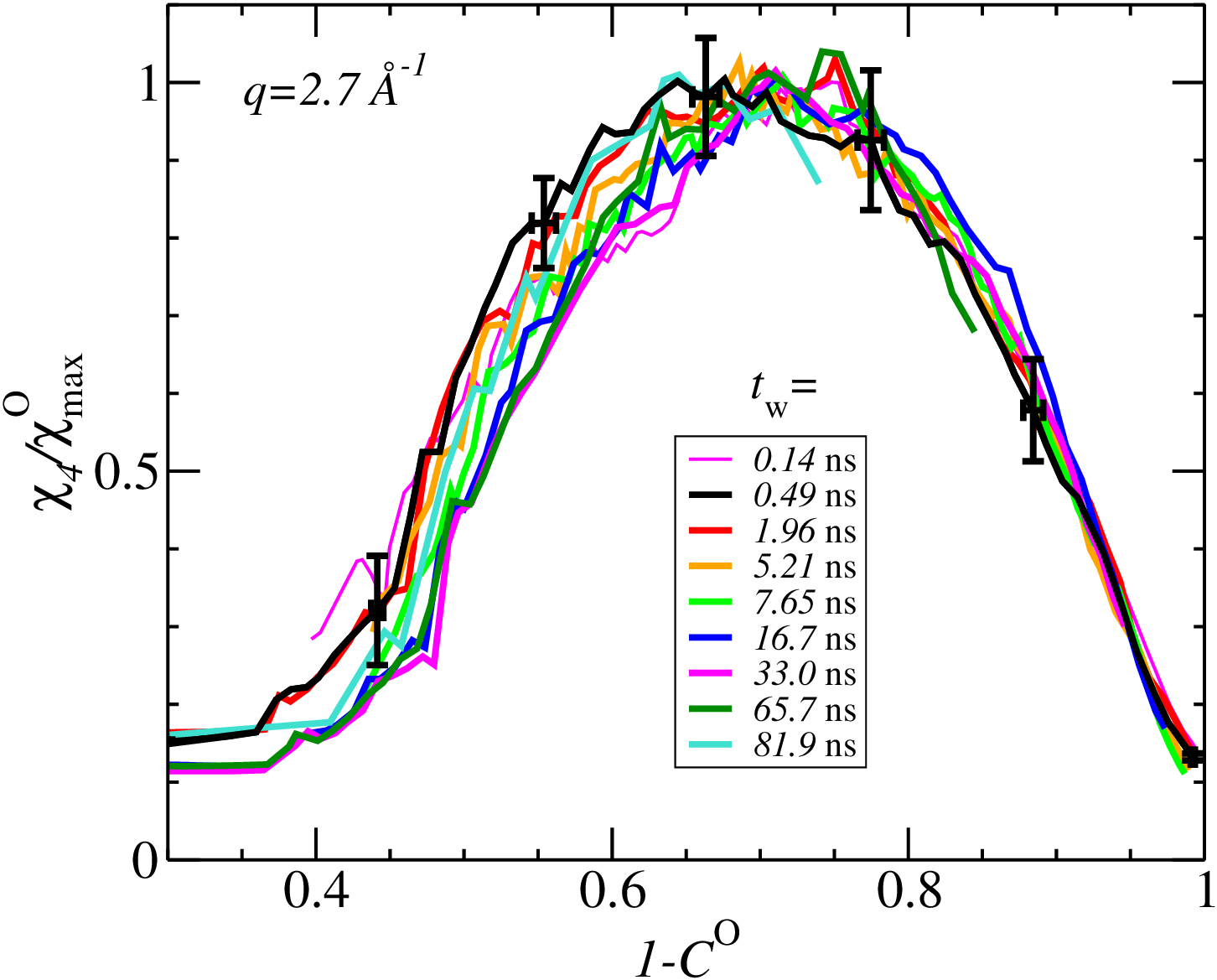}
\caption{\sf (color online) We show here $\chi_4^{\rm O}/\chi_{\rm
    max}^{\rm O}$ as function of $(1-C^{\rm O})$ for different waiting
  times $t_{\rm w}$.  The data collapse confirms
  Eq.~(\ref{eq:chi4horacio}), i.e. that the normalized dynamic
  susceptibility depends on $t_{\rm w}$ only via the global incoherent
  intermediate scattering function $C^{\alpha}$.  We find similar data
  collapse for all other investigated $q$ and $\alpha$.}  
\label{fig:chi4fsscaling_q27_O}
\end{figure}
%-------------
Fig.~\ref{fig:chi4fsscaling_q27_O} shows indeed data collapse 
within the error bars. 
To quantify how good this data collapse is, we determine
the left crossing point of a horizontal line at $0.6$ in 
Fig.~\ref{fig:chi4fsscaling_q27_O}, i.e.
$(1-C^{\alpha})_{\rm cross}$ is defined as the smaller of the two
solutions of the equation

\begin{equation}
\label{eq:Ccrossdef}
\chi_4^{\alpha}/\chi_{\rm max}^{\alpha} ((1-C^{\alpha})_{\rm cross}) = 0.6
\hspace*{1mm}\mbox{.}
\end{equation}

Fig.~\ref{fig:xcrossleft_O} shows the resulting $(1-C^{\alpha})_{\rm
  cross}$ as a function of $t_{\rm w}$ for $\alpha=$O, for $q=1.7
\AA^{-1}$, $2.7 \AA^{-1}$, $3.4\AA^{-1}$, and $4.6\AA^{-1}$.  To a
first approximation, $(1-C^{\alpha})_{\rm cross}$ is waiting time
independent, consistent with Eq.~(\ref{eq:chi4horacio}).  However, a
slight systematic increase of $(1-C^{\alpha})_{\rm cross}$ with
$t_{\rm w}$ appears to be present, which may be evidence for the
existence of small corrections to Eq.~(\ref{eq:chi4horacio}).

%-------------
\begin{figure}
\includegraphics[width=3.4in]{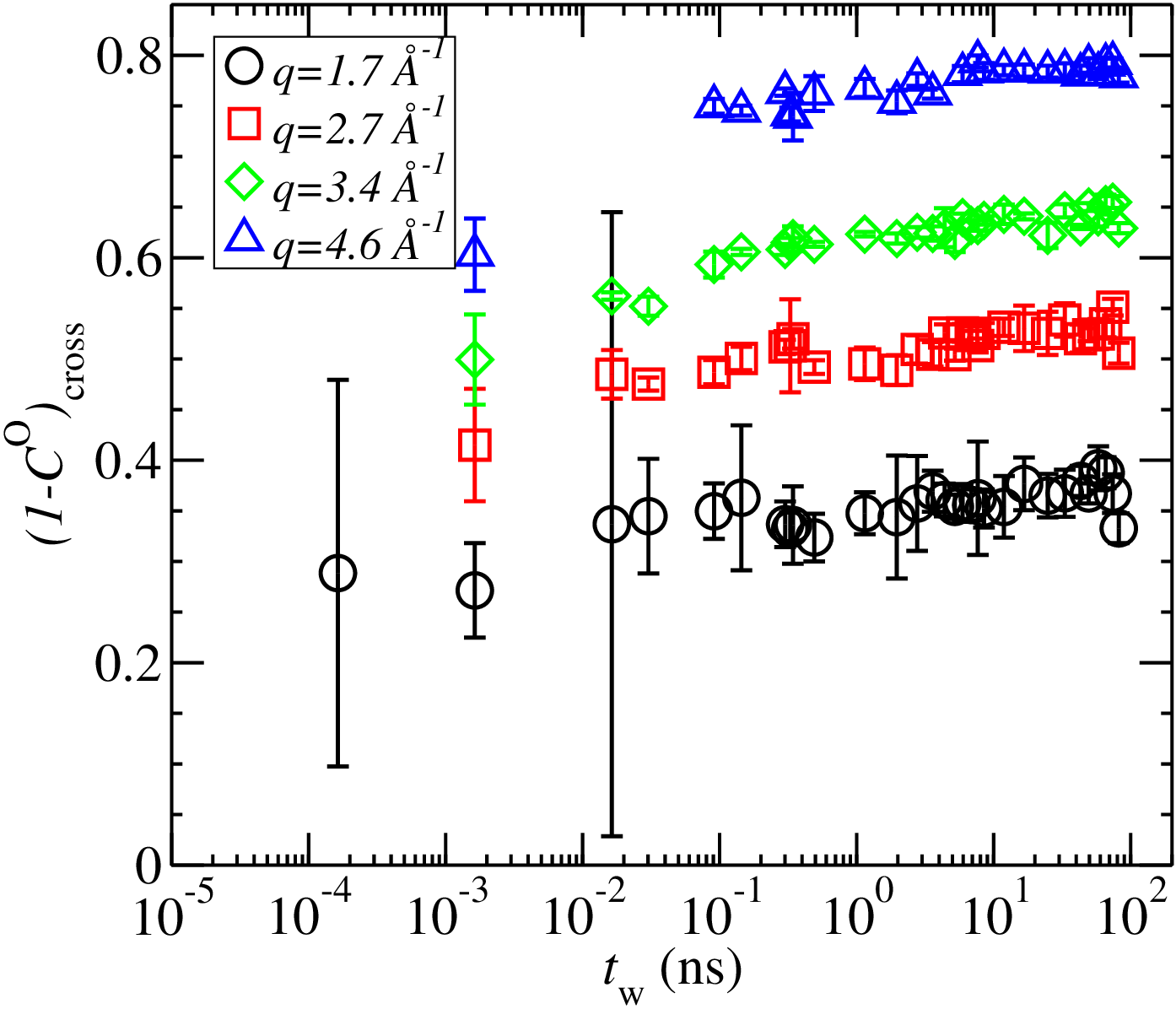}
\caption{\sf (color online) 
   $(1-C^{\rm O})_{\rm cross}$ as defined in 
   Eq.~(\ref{eq:Ccrossdef}) is plotted 
   as function of waiting time $t_{\rm w}$.
}
\label{fig:xcrossleft_O}
\end{figure}
%-------------

%\medskip
We now investigate the dependence of $\chi_4$ on wave vector $q$
and particle type $\alpha$. We use Eq.(\ref{eq:Cqofqalpha}) to
rewrite Eq.(\ref{eq:chi4horacio}) as 

\begin{equation}
\label{eq:chi4qalpha}
\chi_4^{\alpha}(t_{\rm w},t_{\rm w}+t,q)=\chi_{\rm max}(t_{\rm w},q,\alpha) \,
%           \phi^{\alpha}(C(z(t_{\rm w},t),q,\alpha),q)
           \phi(C(z(t_{\rm w},t),q,\alpha),q,\alpha)
\mbox{.}
\end{equation}

The dependence on time $t$ enters in Eq.(\ref{eq:chi4qalpha}) only via
$C$ and therein only through $z(t_{\rm w},t)$, where $z$ is
independent of $q$, as has been shown in Ref.~\cite{kvl2010}, and independent
of $\alpha$, as shown above in
Fig.~\ref{fig:fsOoffsSi_q27q17}. Therefore we have 

\begin{equation}
\label{eq:chi4qalpha_summ}
\chi_4^{\alpha}(t_{\rm w},t_{\rm w}+t,q)=\chi_{\rm max}(t_{\rm w},q,\alpha) \,
           \hat{\chi}(z(t_{\rm w},t),q,\alpha), 
\end{equation}
where $\hat{\chi}(z,q,\alpha)$ is a $t_{\rm w}$-independent function
in the sense that all $(t,t_{\rm w})$-dependence enters only via $z$. 
This means that a parametric plot of $\chi_4^{\alpha}/\chi_{\rm
  max}^{\alpha}$ for $q = q_2$ as a function of
$\chi_4^{\alpha}/\chi_{\rm max}^{\alpha}$ for $q = q_1$, with $q_2
\neq q_1$, should show data collapse. In
Fig.~\ref{fig:chi4ovchi4maxq27ofq17_O}, we show a parametric plot of
this kind, for $q_2=2.7 \AA^{-1}, q_1=1.7 \AA^{-1}$ and $\alpha=$O,
and find that indeed there is data collapse within the error bars. We
also find similar results for other values of $q_2$, $q_1$ and $\alpha$
(not shown).

%-------------
\begin{figure}
\includegraphics[width=3.4in]{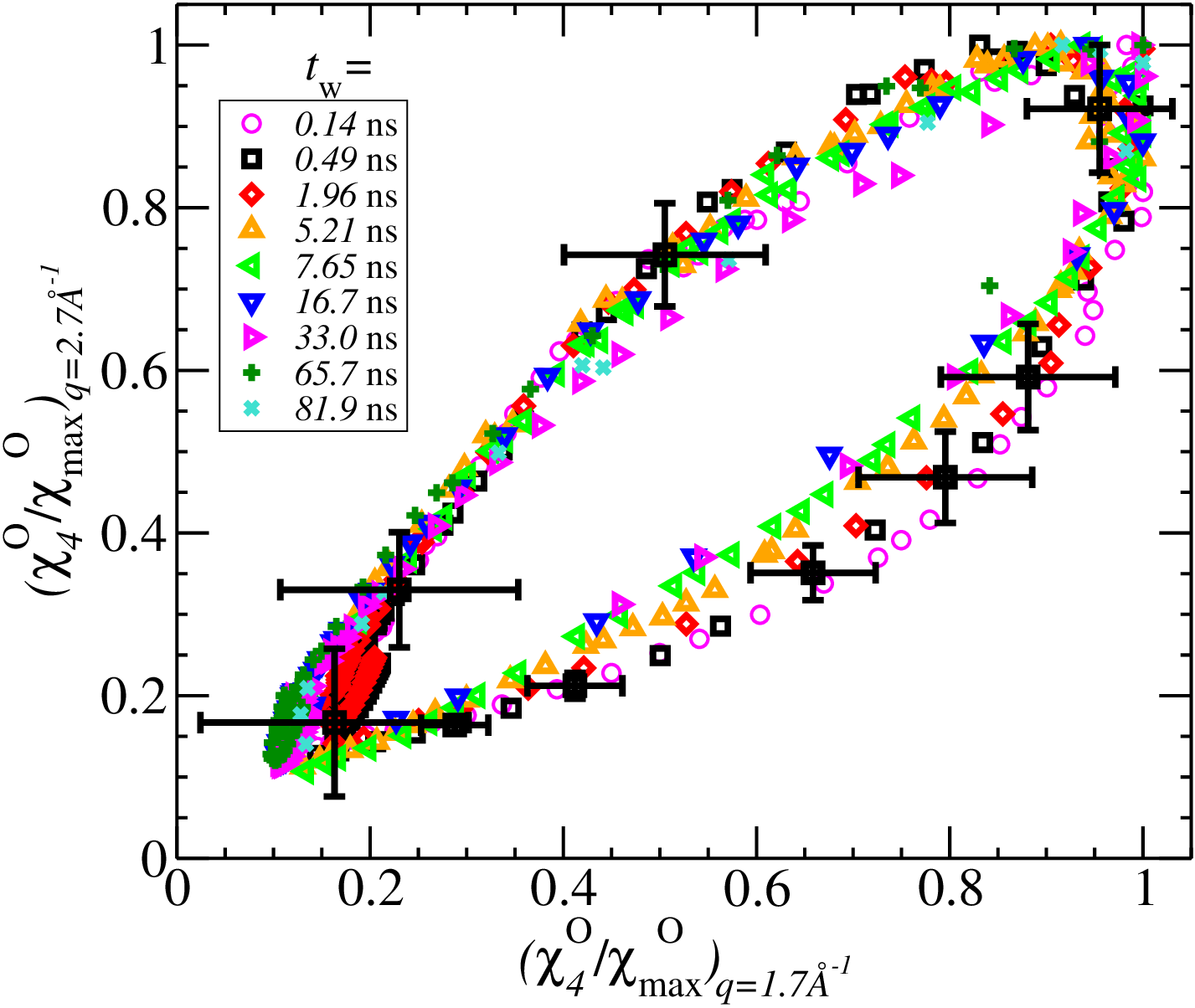}
\caption{\sf (color online) 
   Parametric plot of 
  $\chi_4^{\rm O}/\chi_{\rm max}^{\rm O}(t_{\rm w},t_{\rm w}+t,q=2.7 \AA^{-1})$ 
   versus
  $\chi_4^{\rm O}/\chi_{\rm max}^{\rm O}(t_{\rm w},t_{\rm w}+t,q=1.7 \AA^{-1})$.
  We find data collapse among the results for different waiting times. 
}
\label{fig:chi4ovchi4maxq27ofq17_O}
\end{figure}
%-------------

%\medskip
For the dependence on particle type, $\alpha$,
Eq.~\ref{eq:chi4qalpha_summ} predicts that there should also be data
collapse in a parametric plot of $\large (\chi_4^{\rm O}/\chi_{\rm
  max}^{\rm O} \large )$ versus $\large ( \chi_4^{\rm Si}/\chi_{\rm
  max}^{\rm Si} \large )$.  This data collapse is confirmed with
Fig.~\ref{fig:chi4ovchi4maxOofSi_q27} for $q=2.7 \AA^{-1}$.  We find
equally good collapse for all other investigated $q$ values.  We
emphasize that the data collapse shown in
Figs.~\ref{fig:chi4ovchi4maxq27ofq17_O} and
\ref{fig:chi4ovchi4maxOofSi_q27} is non-trivial, in the sense that 
%$\large (\chi_4^{\rm O}/\chi_{\rm max}^{\rm O} \large )$ versus
%$\large ( \chi_4^{\rm Si}/\chi_{\rm max}^{\rm Si} \large )$ is not
the data are not 
along the diagonal, implying that the shape of
$\chi_4^{\alpha}/\chi_{\rm max}^{\alpha} \,(1-C^{\alpha})$ does depend
both on $\alpha$ and $q$.  However, $z(t_{\rm w},t)$ is independent
of $q$ and $\alpha$ and therefore the $(t_{\rm w},t)$-dependence of
$\chi_4$ is uniquely specified via $C^{\alpha}$, which is a function of
$z(t_{\rm w},t)$.

%-------------
\begin{figure}
\includegraphics[width=3.4in]{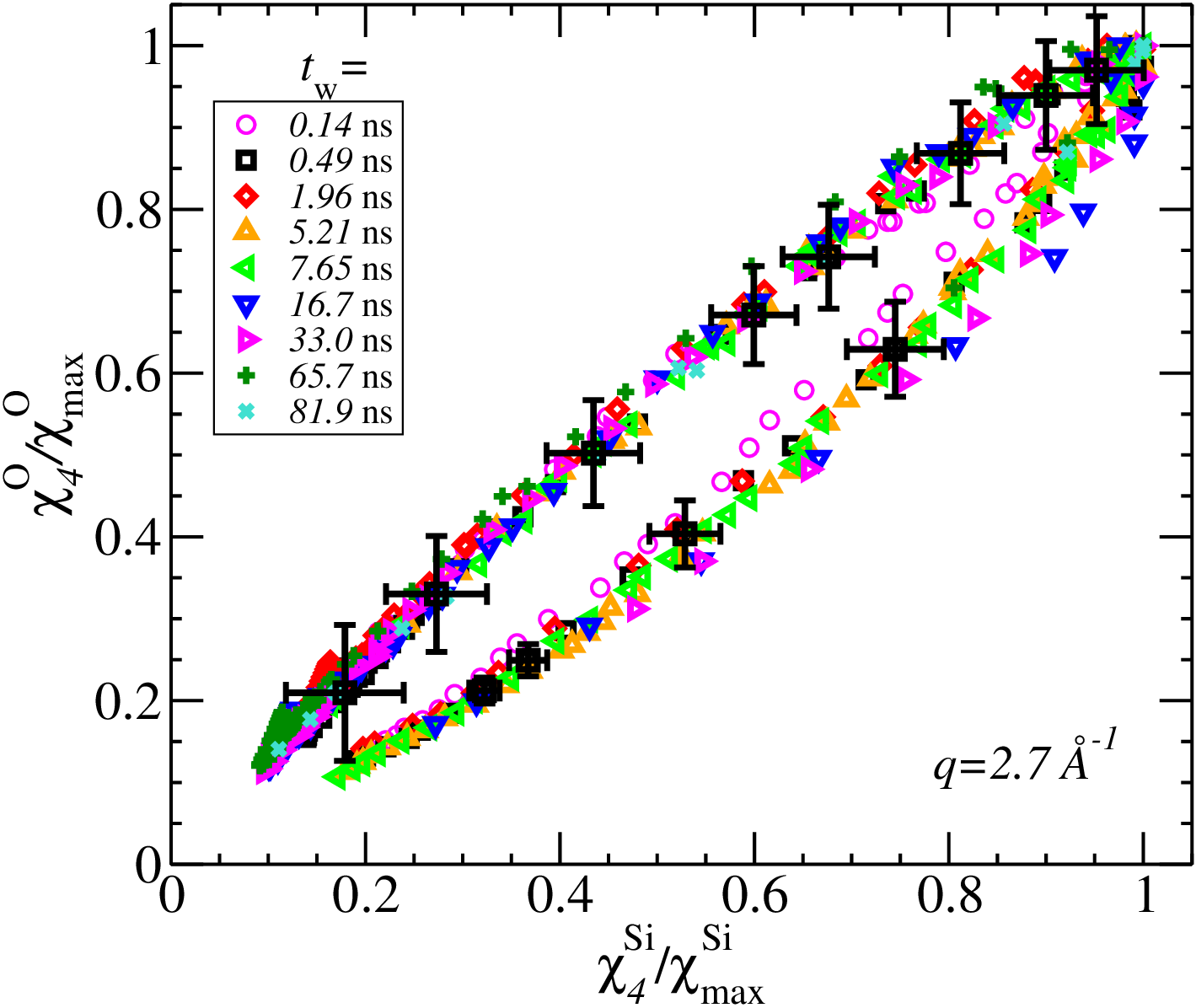}
\caption{\sf (color online) 
   To test the dependence on particle type $\alpha$, this figure shows
   the parametric plot of $\chi_4^{\rm O}/\chi_{\rm max}^{\rm O}$ 
   as function of $\chi_4^{\rm Si}/\chi_{\rm max}^{\rm
     Si}$, for $q=2.7 \AA^{-1}$. The results for different waiting
   times collapse with each other.} 
%   Horacio gave argument that this figure follows from 
%   Fig.~\ref{fig:chi4fsscaling_q27_O} and
%   Fig.~\ref{fig:chi4fsscaling_q27_SiO}, though I am not sure, because
%   chi4ovchi4max is not linear in $alpha$.
\label{fig:chi4ovchi4maxOofSi_q27}
\end{figure}
%-------------

The common aging clock for Si and O atoms allows for the analysis of
Si and O atoms together, thus leading to the data collapse of
$\chi_4^{\textrm{all}}/\chi_{\rm max}^{\textrm{all}}
\,(1-C^{\textrm{all}})$ for different $t_{\rm w}$, which is shown in
Fig.~\ref{fig:chi4fsscaling_q27_SiO} for $q=2.7 \AA^{-1}$ and
quantified for all investigated $q$ values via
$(1-C^{\textrm{all}})_{\rm cross}$ in the inset of
Fig.~\ref{fig:chi4fsscaling_q27_SiO}.  This common aging clock
might be the reason why a data collapse was also found in previous
work on fragile glass formers, in which case different particle types
were analyzed together
\cite{parsaeian08,parsaeian09,parsaeianarXiv2008}.

\begin{figure}
\includegraphics[width=3.4in]{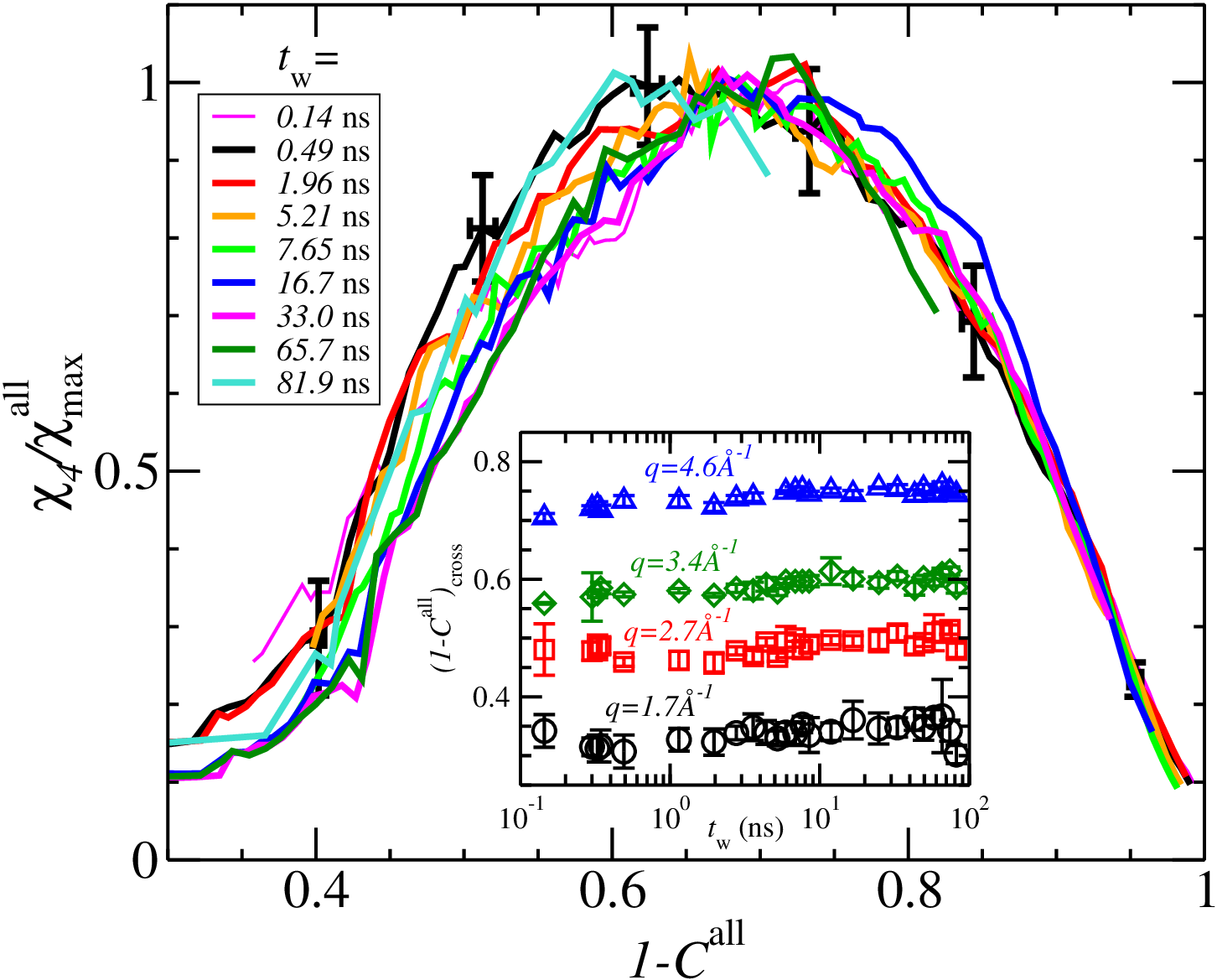}
\caption{\sf (color online) 
   Similar to Fig.~\ref{fig:chi4fsscaling_q27_O} we show here
   $\chi_4^{\alpha}/\chi_{\rm max}^{\alpha}$ as function
   of $(1-C^{\alpha})$ for different waiting times $t_{\rm w}$ but 
   now for $\alpha=$all, i.e. both Si- and O-atoms were included in 
   the analysis. The inset shows the corresponding 
   $(1-C^{\textrm{all}})_{\rm cross}$ as function of $t_{\rm w}$ for all
   investigated $q$ values. }
\label{fig:chi4fsscaling_q27_SiO}
\end{figure}
%-------------

\section{Distribution of Local Incoherent Intermediate Scattering Function}
\label{sec:pofcq}

In the previous section we found scaling for the dynamic
susceptibility, which can be thought of as a measure of the thermal
fluctuations of the {\em global\/} incoherent intermediate scattering
function $f_s^{\alpha}(t_{\rm w},t_{\rm w}+t,{\mathbf q})$.  In this
section we present results on the probability distribution for the
{\em local\/} coarse grained intermediate scattering function
\begin{eqnarray}
\label{eq:Cqlocal}
&& \lefteqn{f_{s,{\mathbf r}}^{\alpha}(t_{\rm w},t_{\rm w}+t,{\mathbf
      q})) =} \\\nonumber 
&& \qquad \qquad \qquad  \frac{1}{N_{\mathbf r}^{\alpha}}
  \sum \limits_{{\mathbf r}_j(t_{\rm w}) \in B_{\mathbf r}}
  \cos \left ( {\mathbf q} \cdot \left [ {\mathbf r}_j(t_{\rm
      w}+t)-{\mathbf r}_j(t_{\rm w}) \right ] \right ), 
\end{eqnarray}
where the sum is over particles of type $\alpha$ which are at time
$t_{\rm w}$ within a local sub-box $B_{\mathbf r}$. By contrast, in
Eq.~(\ref{eq:fs}) the sum is over all particles in the system. Our
definition of $f_{s,{\mathbf r}}$ is identical to the definition of
$C_{\mathbf r}$
in Refs.~\cite{castilloprl88,chamonprl89,castilloprb68,chamonJCP121,castilloprb78}. We choose a different notation here to emphasize
the fact that $f_{s,{\mathbf r}}$ is {\em not\/} an ensemble-averaged
quantity. By definition $f_{s,{\mathbf r}} = 1$ for $t=0$. Relaxation
in a region corresponds to the decay of the value of $f_{s,{\mathbf
    r}}$ from $1$ to $0$.  Spatial fluctuations of $f_{s,{\mathbf r}}$
quantify dynamical heterogeneities: the ``slow'' regions have values
of $f_{s,{\mathbf r}}$ that remain non-negligible for a
longer time, and ``fast'' regions correspond to local values of
$f_{s,{\mathbf r}}$ that decay more rapidly towards $0$.

%\medskip
In the following we determine the probability distribution $P(f_{{\rm
    s},{\mathbf r}}^{\alpha}(t_{\rm w},t_{\rm w}+t,q))$ of the local
correlations $f_{s,{\mathbf r}}^{\alpha}$.  As described in \S
\ref{sec:introduction}, a Landau-theory approximation for spin glasses
\cite{mavimbelaarXiv2013,mavimbelaJStatMech,castilloprb78,chamonJStatMech2007,chamonJCP121,castilloprb68,chamonprl89,castilloprl88}
predicts for this distribution $P(f_{{\rm s},{\mathbf r}}^{\alpha})$
that all $(t_{\rm w},t)$-dependence is solely governed by
$C^{\alpha}(t_{\rm w},t_{\rm w}+t,q)$. This is rather surprising,
since the prediction is for the full distribution $P(f_{{\rm
    s},{\mathbf r}}^{\alpha})$ of these local fluctuations, yet
$C^{\alpha}(t_{\rm w},t_{\rm w}+t,q)$ is not only a scalar but also a
global quantity which is equal to the average
\begin{equation}
C^{\alpha}(t_{\rm w},t_{\rm w}+t,q)=
    \left \langle f_{\rm s}^{\alpha}(t_{\rm w},t_{\rm w}+t,q) \right \rangle \mbox{.}
\label{eq:Cqfsav}
\end{equation}
The theory therefore predicts
that, if $(t_{\rm w},t)$ pairs are chosen such
that $C^{\alpha}(t_{\rm w},t_{\rm w}+t,q)$ is fixed, the corresponding
$P(f_{{\rm s},{\mathbf r}}^{\alpha})$ should be 
$t_{\rm w}$ independent. This data collapse has been confirmed
for spin glasses \cite{chamonJCP121,castilloprl88} and 
for fragile structural glass formers 
\cite{avilaprl2011,parsaeian09,parsaeianarXiv2008,castilloNatPhys2007}. 
% also this?: avilapre2013
%{\bf Horacio: should PRL107,265702(2011) also be cited here?}
%no, instead in intro
In this section we investigate the scaling of 
$P(f_{{\rm s},{\mathbf r}}^{\alpha}(t_{\rm w},t_{\rm w}+t,q))$ for our SiO$_2$ simulation
data, i.e. for a strong glass former.  

%\medskip
As will be shown below, we find that the goodness of the scaling 
depends on the involved length scales via $q$ and via the chosen size 
of the local sub-box $B_{\mathbf r}$. 
Since the specifics of the analysis influence the chosen length scales, we 
include in the following all necessary details.
Our procedure for 
the choice of sub-box size and the corresponding set of sub-boxes
within the complete simulation box of length 
%$L= 16.920468 \AA$ is as follows:
$L= 16.9205 \AA$ is as follows:
We first divide the simulation box into very small sub-boxes 
of length $L/M$. The length of a sub-box $B_{\mathbf r}$ is then
an integer $b$ times this very small sub-box, i.e. 
$L \times b/M$. The average number of particles in $B_{\mathbf r}$ 
is therefore $\langle N_{\mathbf r}\rangle =N_{\alpha} (b/M)^3$. 
We present in this paper results for $\langle N_{\mathbf r}\rangle \approx 5$
and $\langle N_{\mathbf r}\rangle \approx 40$ 
as listed in Table~\ref{table:NBrav}. 
\begin{table}[h]
\centering% NICHT \begin{center}
\begin{tabular}{|c|c|c|c||c|c|c|} \hline 
  $\alpha$ & $M$ & $b$ & $\langle N_{\mathbf r}\rangle$ 
           & $M$ & $b$ & $\langle N_{\mathbf r}\rangle$  \\ \hline \hline
     Si      &   14  &  5   &  5.1  & 14 & 10 & 40.8 \\ \hline
     O       &   7   &  2   &  5.22 & 7  & 4  & 41.8 \\ \hline
     all &   8   &  2   &  5.25 & 8  & 4  & 42.0 \\ \hline
\end{tabular}
\caption{Table of analyzed $(M,b)$-values which specify the local
         sub-box $B_{\mathbf r}$.}
\label{table:NBrav}
\end{table}
The distribution $P(f_{{\rm s},{\mathbf r}}^{\alpha}(t_{\rm w},t_{\rm
  w}+t,q))$ is then obtained for a specific set of $(t_{\rm
  w},t,q,\alpha)$ and for a specific simulation run via measurements
of $f_{s,{\mathbf r}}^{\alpha}$ for all $M^3$ possible $B_{\mathbf r}$
and for all ${\mathbf q}$-vectors of magnitude $q$.  To obtain the
$M^3$ possible sub-boxes $B_{\mathbf r}$, the sub-box is shifted in the
three directions and periodic boundary conditions were used.

%\medskip
To test whether the $(t_{\rm w},t)$ dependence of $P(f_{{\rm
    s},{\mathbf r}}^{\alpha})$ is governed by $C^{\alpha}(t_{\rm
  w},t_{\rm w}+t,q)$, we use the same approach as in previous work
\cite{avilaprl2011,parsaeian09,parsaeianarXiv2008,castilloNatPhys2007}.
We choose a fixed value $C_{\rm fix}$ of the global $C^{\alpha}$, and
for each waiting time $t_{\rm w}$ we determine the time $t_{\rm fix}$
such that $C^{\alpha}(t_{\rm w},t_{\rm w}+t_{\rm fix},q) = C_{\rm
  fix}$ (see endnote \footnote{\label{Calphanote} A desired $C^{\alpha}$ value can
  be achieved only up to a certain accuracy, because during any
  simulation run configurations are saved only at certain discrete
  times.  Of the available configurations, we choose the one for which
  $C^{\alpha}$ is the closest to $C_{\rm fix}$. Results are reported
  only for cases when $C^{\alpha}$ is within $1\%$ of the chosen
  $C_{\rm fix}$ value.
}.) %footnote
We then determine for each independent simulation run 
$c=1, 2, \ldots, 200$, a distribution 
$P_c(f_{s,{\mathbf r}}^{\alpha}(t_{\rm w},t_{\rm w}+t_{\rm fix},q))$, 
using for Eq.~\ref{eq:Cqlocal} the positions 
$\left\{{\mathbf r}_i(t_{\rm w})\right\}_c$ and 
$\left\{{\mathbf r}_i(t_{\rm w}+t_{\rm fix})\right\}_c$. Please note that 
the parameters $t_{\rm w}$, $t_{\rm w}+t_{\rm fix}$ and $q$ 
are common to all simulation runs.  
For each bin of the distribution, we obtain the mean $P_c(f_{s,{\mathbf
    r}}^{\alpha}(t_{\rm w},t_{\rm w}+t_{\rm fix},q))$ over the 200
independent simulation runs. The error bars represent the standard
deviation of this mean over runs. The result of this computation is shown in Fig.~\ref{fig:pofcq_q27_Cqfix050_O_b2}
%and \ref{fig:pofcq_q34_Cqfix040_O_b2} 
for the case of $\alpha =$O and $\langle N_{\mathbf r}\rangle=5.1$.

%-------------
\begin{figure}
\includegraphics[width=3.4in]{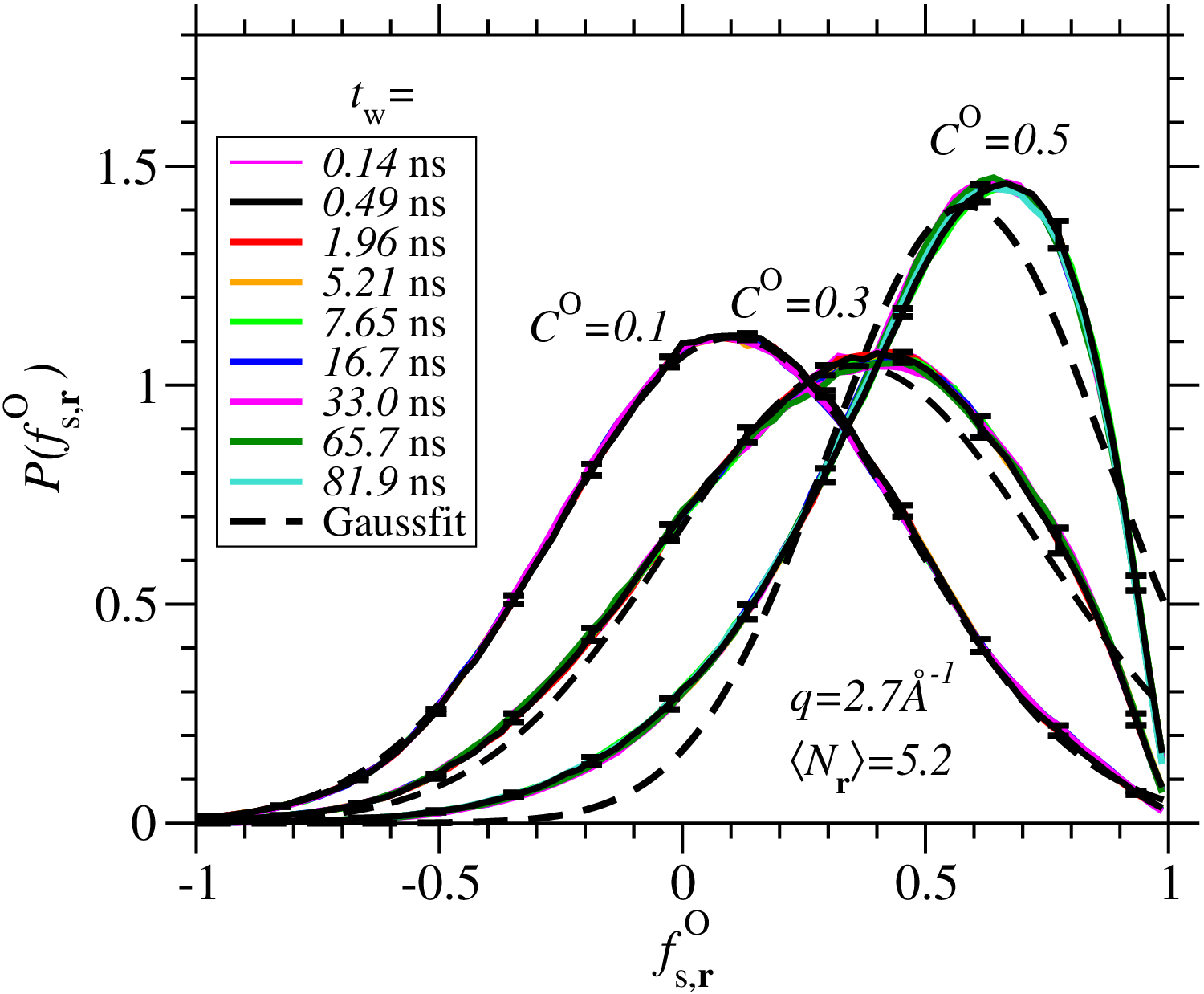}
\caption{\sf (color online) Distribution of the local
   incoherent intermediate scattering function  for oxygen atoms,
   $P(f_{{\rm s},{\mathbf r}}^{\rm O})$, for $q=2.7 \AA^{-1}$ and 
   $\langle N_{\mathbf r}\rangle=5.1$.
}
\label{fig:pofcq_q27_Cqfix050_O_b2}
\end{figure}
%-------------

For small $C^{\alpha}$ we find perfect scaling collapse and the
distribution is a Gaussian (black dashed line). We attribute the
Gaussian distribution to small $C^{\alpha}$ occurring at late times
$t$ (see Fig.~\ref{fig:fslntav_q27_O}) when diffusive dynamics is
approached.  For intermediate and large $C^{\alpha}$ we find a
non-Gaussian, i.e. non-trivial $P(f_{{\rm s},{\mathbf r}})$, and
nevertheless very good data collapse. To check quantitatively whether
the slight $t_{\rm w}$ dependence in
Figs.~\ref{fig:pofcq_q27_Cqfix050_O_b2} is systematic, we focus on the
location of most discrepancy, the maximum.
Fig.~\ref{fig:heightmaxPofCq_O_b2_q27} shows the maximum value
$P^{\alpha}_{\rm max}$ of the distribution, as a function of $t_{\rm
  w}$. Consistent with the results above, we find that for small
$t_{\rm w}$ scaling is not valid, but for $t_{\rm w} \gtrsim 0.1$~ns
$P^{\alpha}_{\rm max}$ is approximately independent of $t_{\rm w}$.
We obtain similar results for $q>2.7 \AA^{-1}$ and also for
$\alpha=$Si.

%-------------
\begin{figure}
\includegraphics[width=3.4in]{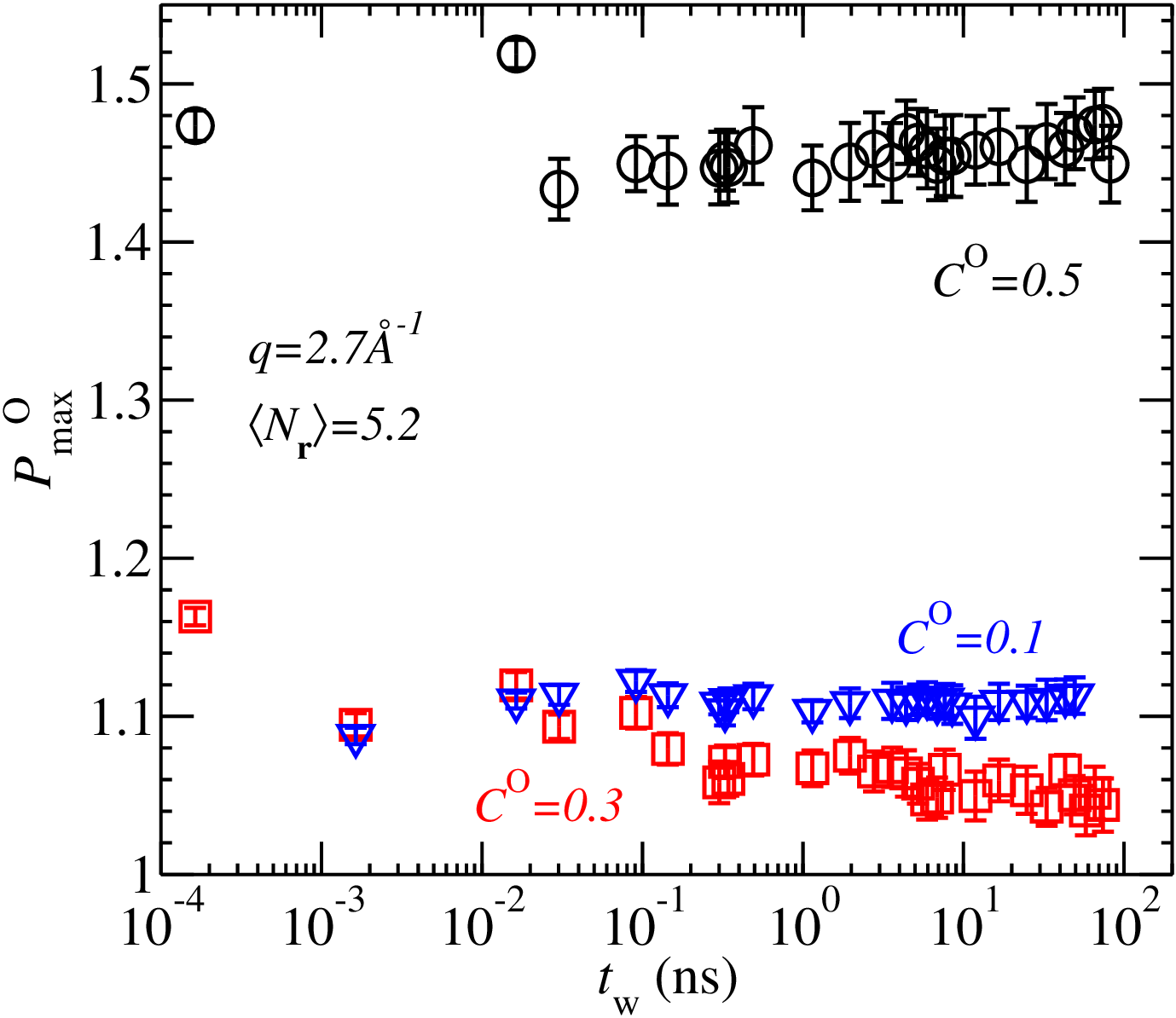}
\caption{\sf (color online) 
   To quantify how well scaling is satisfied in
   Fig.\ref{fig:pofcq_q27_Cqfix050_O_b2}, 
   we show here the maximum $P^{\rm O}_{\rm max}$ as a function of the 
   waiting time $t_{\rm w}$. Here $\langle N_{\mathbf r} \rangle=5.1$ and 
   $q=2.7 \AA^{-1}$. We find similar 
   results for $q > 2.7 \AA^{-1}$.
}
\label{fig:heightmaxPofCq_O_b2_q27}
\end{figure}
%-------------

To probe the dependence of $P(f_{{\rm s},{\mathbf r}}^{\alpha})$ on particle
type $\alpha$, we show in the inset of 
Fig.~\ref{fig:pofcq_q27_Cqfix050_030_SiO_b2_insetalpha}
the comparison of $P(f_{{\rm s},{\mathbf r}}^{\alpha})$ for
$\alpha=$Si, O, all. 
We conclude that $P(f_{{\rm s},{\mathbf r}}^{\alpha})$ does depend on $\alpha$.
In previous sections \S \ref{sec:Fs} and \S \ref{sec:chi4} we had 
found that despite different dynamics of silicon and oxygen atoms, 
their scaling $z(t_{\rm w},t)$ gives rise to a common aging clock.
This common clock allows us to analyze Si and O together 
($\alpha=$all).
To test this common aging clock for the case of
$P(f_{{\rm s},{\mathbf r}}^{\alpha})$, we therefore show in 
Fig.~\ref{fig:pofcq_q27_Cqfix050_030_SiO_b2_insetalpha}
$P(f_{{\rm s},{\mathbf r}}^{\textrm{all}})$ and find 
indeed a data collapse for different $t_{\rm w}$.

\begin{figure}
\includegraphics[width=3.4in]{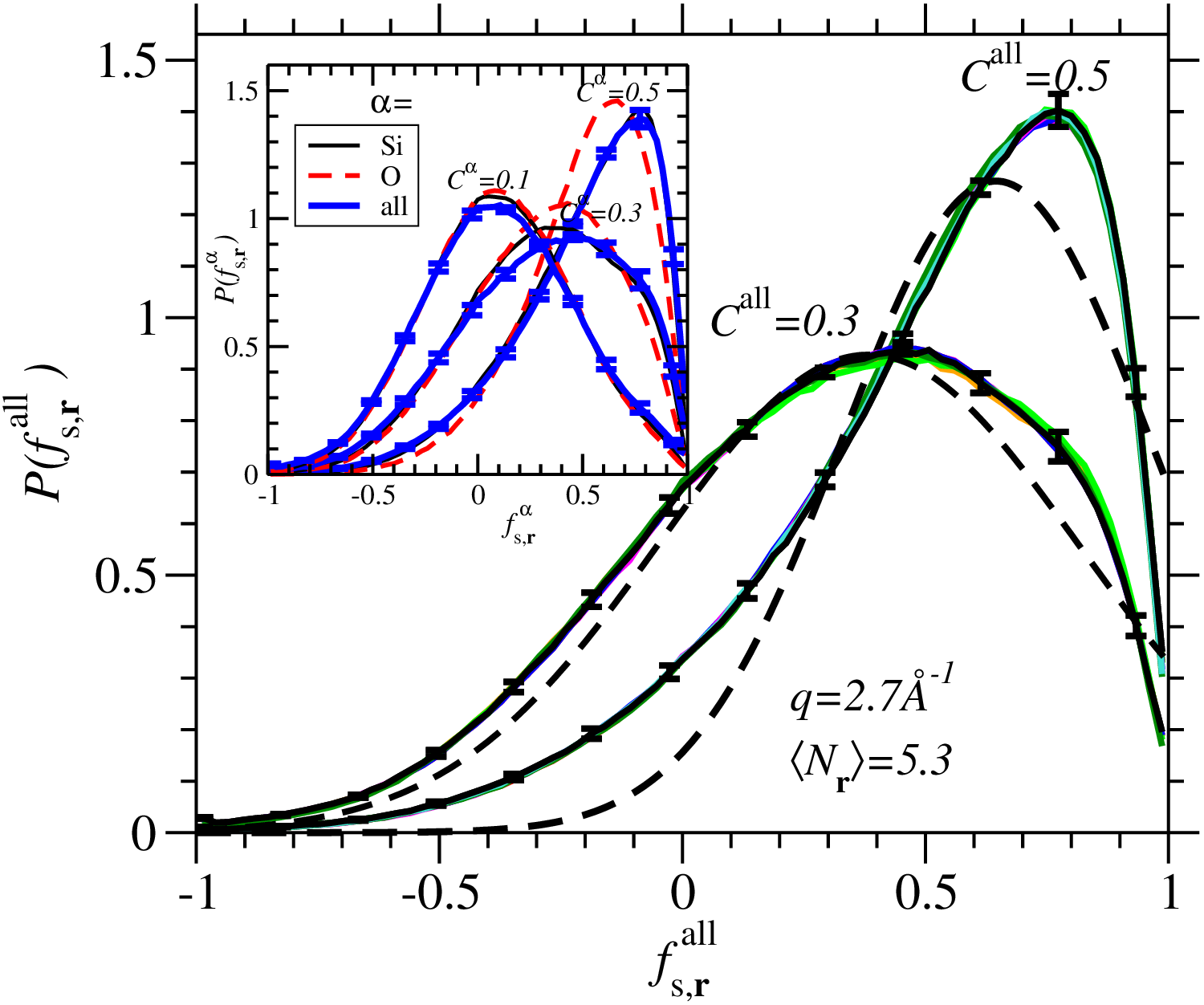}
\caption{\sf (color online) 
    Distribution of the local incoherent intermediate scattering 
    function for $\alpha=$all, i.e. when both silicon and oxygen 
    atoms are analyzed. The color coding for the waiting times is the same
    as in Fig.~\ref{fig:pofcq_q27_Cqfix050_O_b2}. 
    The dashed black lines correspond to Gaussian fits. For
    comparison, the inset shows 
    $P(f_{{\rm s},{\mathbf r}}^{\alpha})$ for different $\alpha$
    with fixed waiting time $t_{\rm w}=16.7$~ns.
}
\label{fig:pofcq_q27_Cqfix050_030_SiO_b2_insetalpha}
\end{figure}
%-------------

%\medskip
So far we have shown $P(f_{{\rm s},{\mathbf r}}^{\alpha})$ for various
fixed $C^{\alpha}$, for $q=2.7$ \AA$^{-1}$, $\langle N_{\mathbf
  r}\rangle \approx 5$, with either $\alpha=$O or $\alpha=$all.  When
also $q$ and $\langle N_{\mathbf r}\rangle$ are varied, we find that
the predicted data collapse occurs as long as $t_{\rm w} \gtrsim
0.1$~ns, $q \geq 2.7$ \AA$^{-1}$ and $\langle N_{\mathbf r}\rangle
\approx 5$.  Next we look at cases when scaling fails.
Fig.~\ref{fig:pofcq_q17_Cqfix050_030_O_b2_insetPmax} shows $P(f_{{\rm
    s},{\mathbf r}}^{\alpha})$ for $\langle N_{\mathbf r} \rangle
\approx 5$ as before, but now $q=1.7 \AA^{-1}$.
\begin{figure}[h]
\includegraphics[width=3.4in]{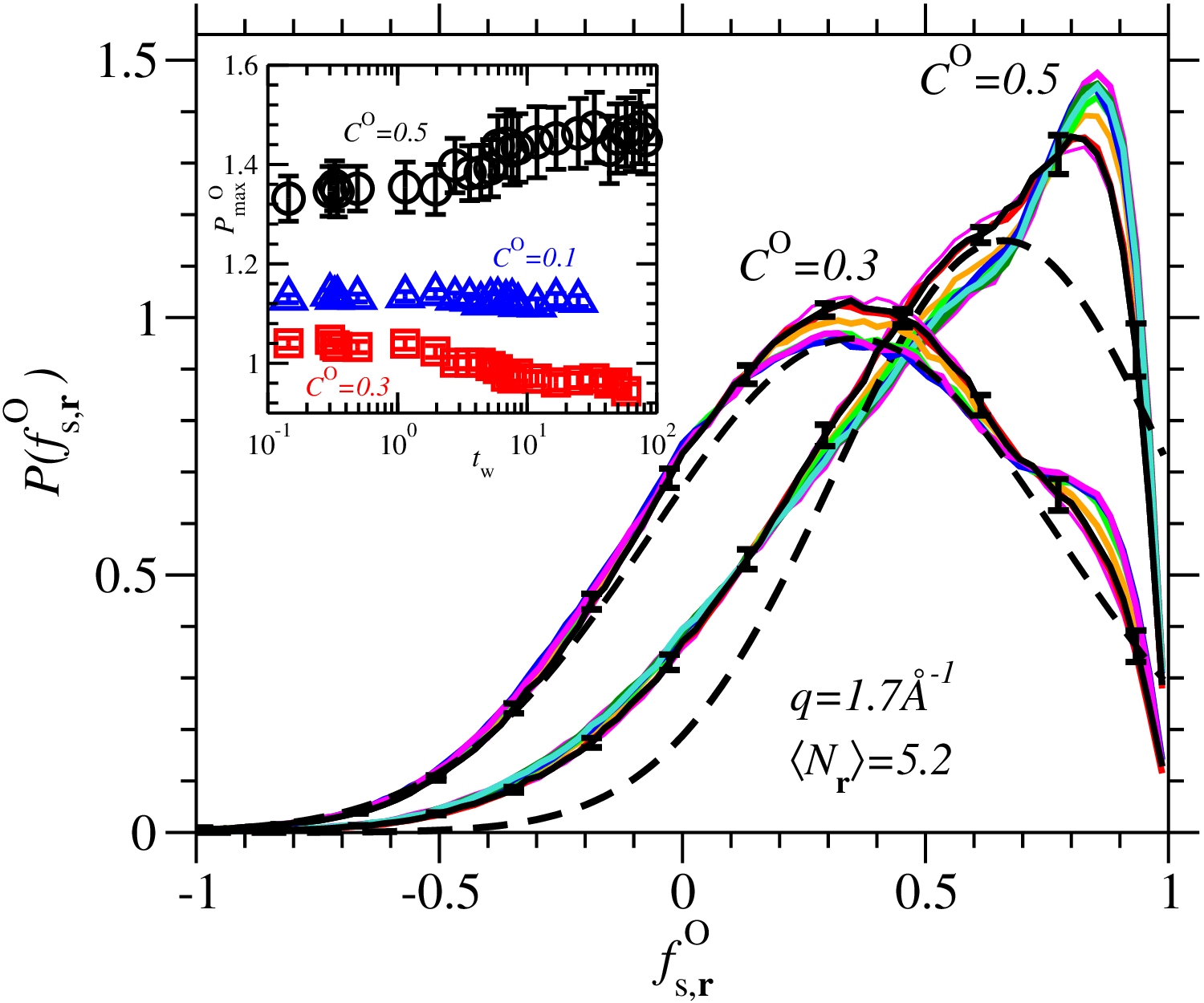}
\caption{\sf (color online) 
   $P(f_{{\rm s},{\mathbf r}}^{\rm O})$ for $q=1.7 \AA^{-1}$. Scaling
   is less good when $q$ is too small. The inset shows 
   the maximum value, $P_{\rm max}^{\rm O}$, 
   as a function of $t_{\rm w}$. Even for $t_{\rm w} > 0.1$,
   $P_{\rm max}^{\rm O}$ is $t_{\rm w}$-dependent, which indicates
   that there is a breakdown of scaling.
}
\label{fig:pofcq_q17_Cqfix050_030_O_b2_insetPmax}
\end{figure}
%-------------
%
For $C^{\rm O}=0.3$ and $C^{\rm O}=0.5$ we find that the data collapse
for different $t_{\rm w}$ is much worse than before. This is
quantified in the inset of
Fig.~\ref{fig:pofcq_q17_Cqfix050_030_O_b2_insetPmax}, which shows the
systematic $t_{\rm w}$ dependence of $P_{\rm max}^{\rm O}$.  The inset
also shows that in the Gaussian case of $C^{\rm O}=0.1$, the scaling
does work even for $q=1.7 \AA^{-1}$.  Furthermore, when the sub-box
size is chosen to be much larger, scaling does not occur even if $q
\ge 2.7 \AA^{-1}$, as shown in
Fig.~\ref{pofcq_q27_Cqfix050_030_O_b4_insetPmax} for the case of
$\langle N_{\mathbf r} \rangle \approx 42$.
\begin{figure}[h]
\includegraphics[width=3.4in]{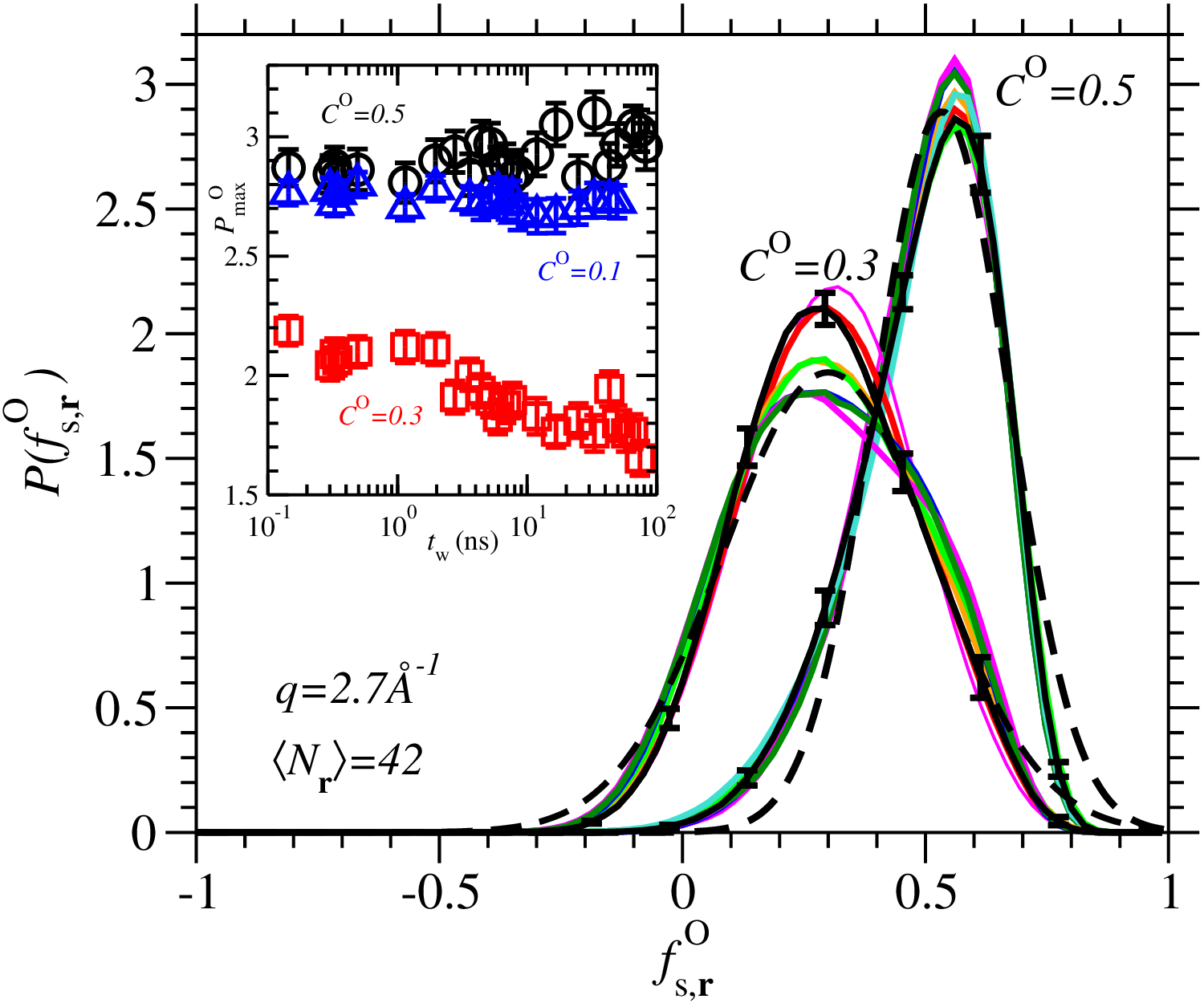}
\caption{\sf (color online) $P(f_{{\rm s},{\mathbf r}}^{\rm O})$ for
  sufficiently large $q=2.7 \AA^{-1}$ but with a large sub-box size,
  $\langle N_{\mathbf r} \rangle=41.8$, leading to $t_{\rm
    w}$-dependent distributions. To quantify the breakdown of scaling
  the inset shows the peak value $P_{\rm max}^{\rm O}$ as a function of
  the waiting time $t_{\rm w}$.  }
\label{pofcq_q27_Cqfix050_030_O_b4_insetPmax}
\end{figure}
%-------------

%\medskip
Thus scaling breaks down for intermediate timescales, corresponding to
intermediate $C^{\alpha}$, when the regions probed by the local
incoherent intermediate scattering function $f_{{\rm s},{\mathbf r}}$
become too large. Those probed regions can become larger either
directly, because the coarse graining region is chosen to be larger,
or indirectly, because $q$ is chosen to be smaller, thus allowing
longer displacements to contribute significantly to $f_{{\rm
    s},{\mathbf r}}$. 

One effect that could contribute to the imperfect collapse is that the
measured probability distribution could be influenced to some degree
by spatial correlation effects, and these effects could vary as the
system ages and the typical size of the correlated regions grows. In
the case of fragile glasses~\cite{castilloNatPhys2007}, it was found
that the width of $P(f_{{\rm s},{\mathbf r}})$ grew with $t_{\rm w}$
at constant $C^{\alpha}$. 
In that case it was argued that $P(f_{{\rm s},{\mathbf r}})$ was narrowed by
averaging of $f_{{\rm s},{\mathbf r}}$ over more than one correlated
region, but this narrowing became weaker as the size of the correlated
regions grew with $t_{\rm w}$. In our case, however, we notice in the
insets of Figs.~\ref{fig:pofcq_q17_Cqfix050_030_O_b2_insetPmax} and
\ref{pofcq_q27_Cqfix050_030_O_b4_insetPmax} that the direction of the
effect is not always the same: the distributions {\em widen\/} with
increasing $t_{\rm w}$ for $C^{O}=0.3$, but they {\em narrow\/} with
increasing $t_{\rm w}$ for $C^{O}=0.5$. Therefore, for SiO$_2$,
although this narrowing effect could in principle play some role,
there must also be other effects at play.

In what follows we address the question of why scaling fails and how
to adjust the analysis to recover the data collapse even for larger
length scales ($q=1.7$~\AA$^{-1}$ and $\langle N_{\mathbf r} \rangle
\approx 40$). To gain this insight, first a closer look at the details
of the analysis is necessary. A crucial point is how we choose the
time pairs, i.e.  $t_{\rm w}$ and $t_{\rm w}+t$. We illustrate in
Fig.~\ref{fig:fsoft_q17_O_tw100000_c0127c0057c0007c0121} how this is
done. First we choose a unique $t=t_{\rm fix}$ --- the same for all
$200$ independent simulation runs --- by demanding that the global
incoherent scattering function $C^{\alpha}(t_{\rm w}, t_{\rm w}+t)$
(thick line in
Fig.~\ref{fig:fsoft_q17_O_tw100000_c0127c0057c0007c0121}) take a
certain value at $t=t_{\rm fix}$, such as $C^{\alpha}(t_{\rm w},
t_{\rm w}+t_{\rm fix}) =0.5$. 

If we now look at individual runs, the
intermediate scattering function computed for each run $c$ is
$C_c^{\alpha}=\langle f_{\rm s}\rangle_c$ where $\langle \ldots
\rangle_c$ corresponds to an average over ${\mathbf q}$ vectors of
fixed magnitude but not over simulation runs. Since the system
simulated in each independent run contains only 336 particles, it is
not large enough for $C_c^{\alpha}$ to be self-averaging: in
Fig.~\ref{fig:fsoft_q17_O_tw100000_c0127c0057c0007c0121}, the values
of $C_c^{\alpha}(t_{\rm w}, t_{\rm w}+t_{\rm fix})$ for four
individual runs are shown with circles, and they differ dramatically
from each other and from the value of the fully averaged
$C^{\alpha}(t_{\rm w}, t_{\rm w}+t_{\rm fix})$. In other words,
choosing a unique value $t=t_{\rm fix}$ is equivalent to choosing very
different values for $C_c^{\alpha}$ for each simulation run
$c$. Consequently, the distributions $P(f_{{\rm s},{\mathbf r}}(t_{\rm
  w},t_{\rm w}+t,q))$ at the same times are necessarily very different
for different runs $c$. 

In Fig.~\ref{fig:pofcq_q17_Cqfix050_simruns_O_b2}, single simulation run
distributions $P_c(f_{{\rm s},{\mathbf r}}^{\rm O})$ for four
independent runs are shown as thin black lines, and it is clear that
the variation of $P_c(f_{{\rm s},{\mathbf r}}^{\rm O})$ between runs
is very large: there is a nontrivial {\em distribution of
  distributions\/}. For comparison, the average over runs $P(f_{{\rm
    s},{\mathbf r}}^{\rm O})$ is shown in the same figure with a thick
black line. All of the black lines in the figure correspond to $t_{\rm
  w}=0.49$~ns. In the same figure, results are shown for another
waiting time, $t_{\rm w}=33.0$~ns, as magenta/grey lines: the thin
lines corresponding to $P_c(f_{{\rm s},{\mathbf r}}^{\rm O})$ for
individual runs $c$, and the thick line corresponding to $P(f_{{\rm
    s},{\mathbf r}}^{\rm O})$.
%-------------
\begin{figure}[h]
\includegraphics[width=3.4in]{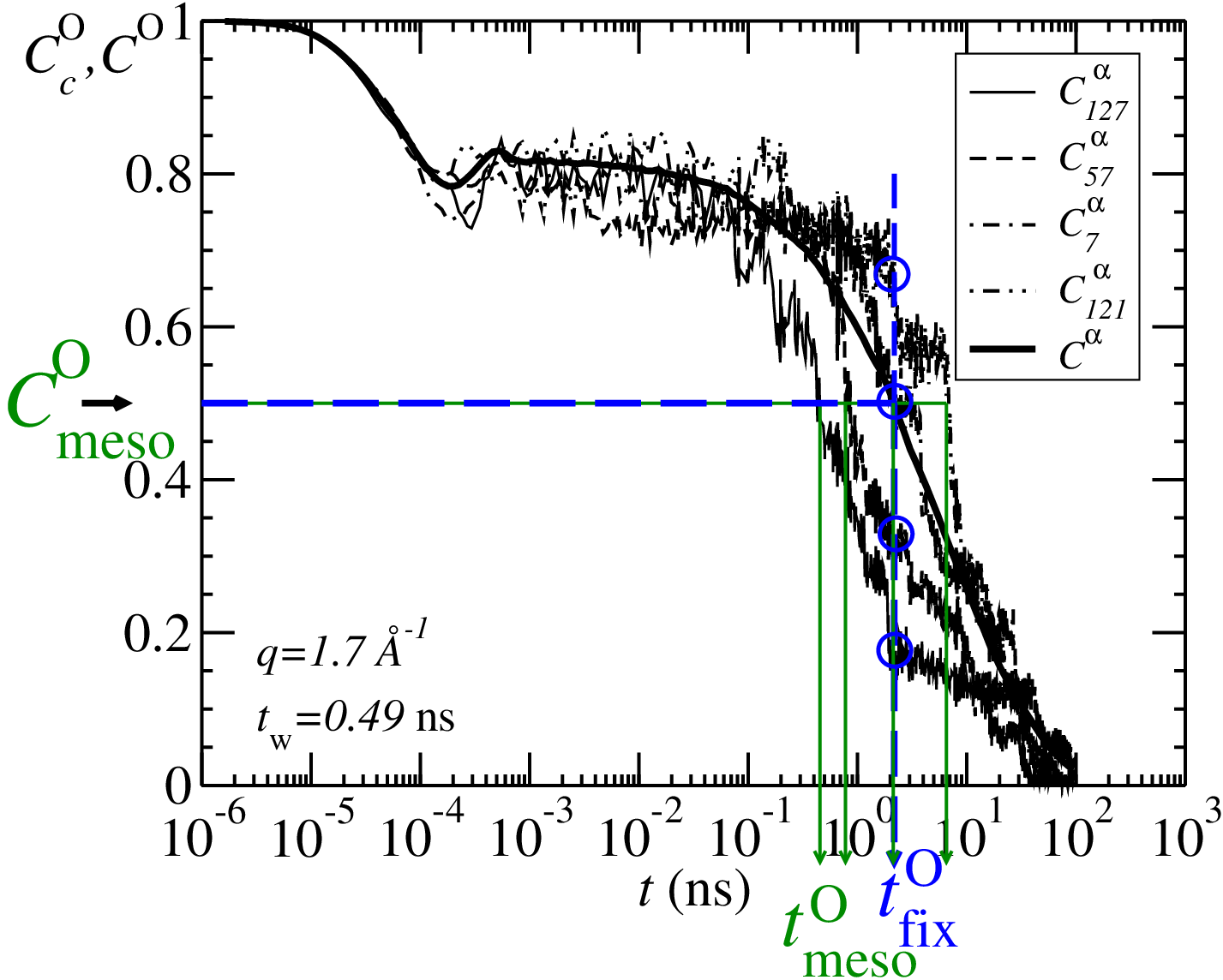}
\caption{\sf (color online) 
  $C^{\alpha}_c$ for single simulation runs $c=127, 57, 7,$ and $121$ 
  (thin lines)
  and the averaged incoherent intermediate scattering function $C^{\alpha}$
  (thick line).
  In this example $\alpha = $O.
}
\label{fig:fsoft_q17_O_tw100000_c0127c0057c0007c0121}
\end{figure}
%-------------
\begin{figure}[h]
\includegraphics[width=3.4in]{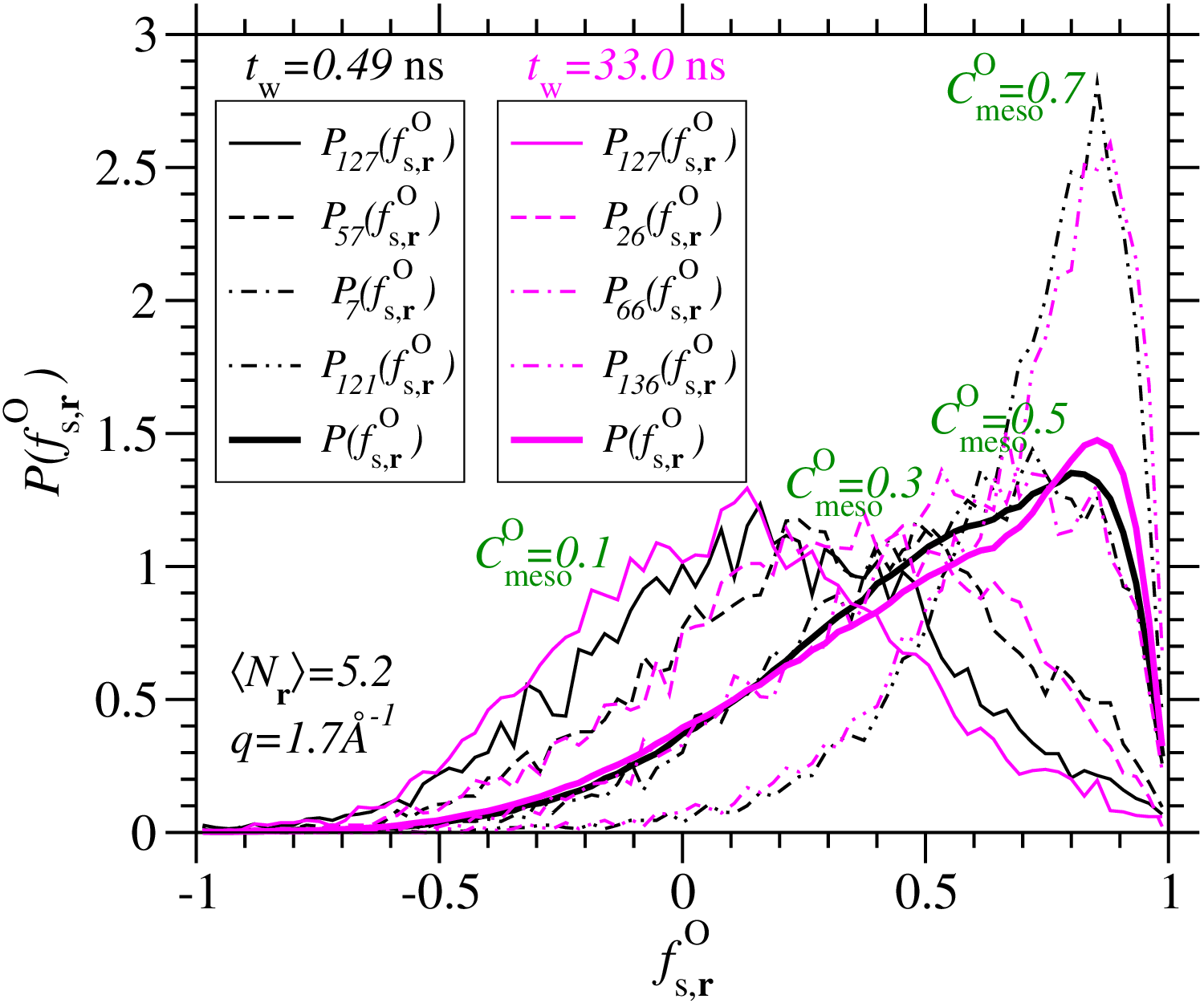}
\caption{\sf (color online) 
   Distributions of local incoherent intermediate scattering
   functions.  
   Thick lines correspond to the average distribution
   $P(f_{{\rm s},{\mathbf r}})$ and thin lines are examples 
   for individual simulation run distributions $P_c(f_{s,{\mathbf r}})$.
   Colors (black and magenta) indicate the waiting times.
}
\label{fig:pofcq_q17_Cqfix050_simruns_O_b2}
\end{figure}
%-------------
%Therefore, by choosing above $t_{\rm fix}$ the same for
%all independent simulation runs, we were assuming that the system 
%is self-averaging, i.e. we introduced indirectly a length scale,
%over which we averaged. This assumption is working suprisingly
%well, given how similar the averages in 
%Fig.~\ref{fig:pofcq_q17_Cqfix050_simruns_O_b2} (bold lines) are 
%compared to the large fluctuations in the individual distributions
%for each simulation run. 
Also, for $t_{\rm w}=33.0$~ns the distributions $P_c(f_{{\rm
    s},{\mathbf r}}^{\rm O})$ vary greatly from simulation run to
simulation run. Yet, the set of possible $P_c(f_{{\rm s},{\mathbf
    r}}^{\rm O})$ seem to be the same for the two waiting times. We
find that for obtaining a particular shape of $P_c(f_{{\rm s},{\mathbf
    r}}^{\alpha})$, the key variable is $C_c^{\alpha}(t_{\rm fix}^{\rm
  O})$ (marked by circles in
Fig.~\ref{fig:fsoft_q17_O_tw100000_c0127c0057c0007c0121}). For
example, in Fig.~\ref{fig:pofcq_q17_Cqfix050_simruns_O_b2} the two
thin dashed lines correspond to two different simulation runs, $c =
26$ for $t_{\rm w}=33.0$~ns and $c = 57$ for $t_{\rm w}=0.49$~ns,
chosen so that in both cases $C_c^{O}(t_{\rm fix}^{\rm O}) \approx
0.3$. This makes the two distributions close enough that
their differences are of the order of their statistical error. The
figure also shows that the same procedure is successful for obtaining
other pairs of nearly identical distributions for $C_c^{O}(t_{\rm
  fix}^{\rm O})\approx 0.1,0.5,0.7$ 
(for more details see endnote \footnote{To be precise,
for $t_{\rm w}=0.49$~ns the lowest possible $C_c^{\rm O}$ is $0.165$, which
was chosen here. For all other cases shown in 
Fig.~\ref{fig:pofcq_q17_Cqfix050_simruns_O_b2} the target value of $C_c^{\rm
  O}$ is achieved with $1$\% accuracy or better.}.)

%\medskip
This leads us directly to finding a way to improve the scaling even
for longer length scales. We no longer use a unique time $t_{\rm fix}$
which is the same for all simulation runs $c$. Instead, we specify a fixed
value $C_{\rm meso}^{\alpha}$ of the correlation, and we define, for
each simulation run $c$, a time $t^{\alpha}_{\rm meso}(c)$ such that
\begin{equation}
C_{\rm meso}^{\alpha}=
 \left \langle f_{\rm s}^{\alpha}(t_{\rm w},t_{\rm w}+t^{\alpha}_{\rm meso}(c),q)
 \right \rangle_c, 
\label{eq:tmeso}
\end{equation}
as shown in
Fig.~\ref{fig:fsoft_q17_O_tw100000_c0127c0057c0007c0121}. Thus, for a
specified $C_{\rm meso}^{\alpha}$, and for each run $c=1, 2, \ldots,
200$, we determine $t^{\alpha}_{\rm meso}(c)$, the corresponding $P_c(f_{{\rm
    s},{\mathbf r}}^{\alpha}(t_{\rm w},t_{\rm w}+t^{\alpha}_{\rm meso}(c),q))$,
and then we average all of the individual distributions $P_c(f_{{\rm
    s},{\mathbf r}}^{\alpha})$ to obtain $P(f_{{\rm s},{\mathbf
    r}}^{\alpha})$ (see endnote \footnote{In our analysis we 
only use simulation
  runs $c$ for which there exist times such that Eq.~\ref{eq:tmeso} is
  satisfied within an accuracy of 1\% or better.}.)
\begin{figure}
\includegraphics[width=3.4in]{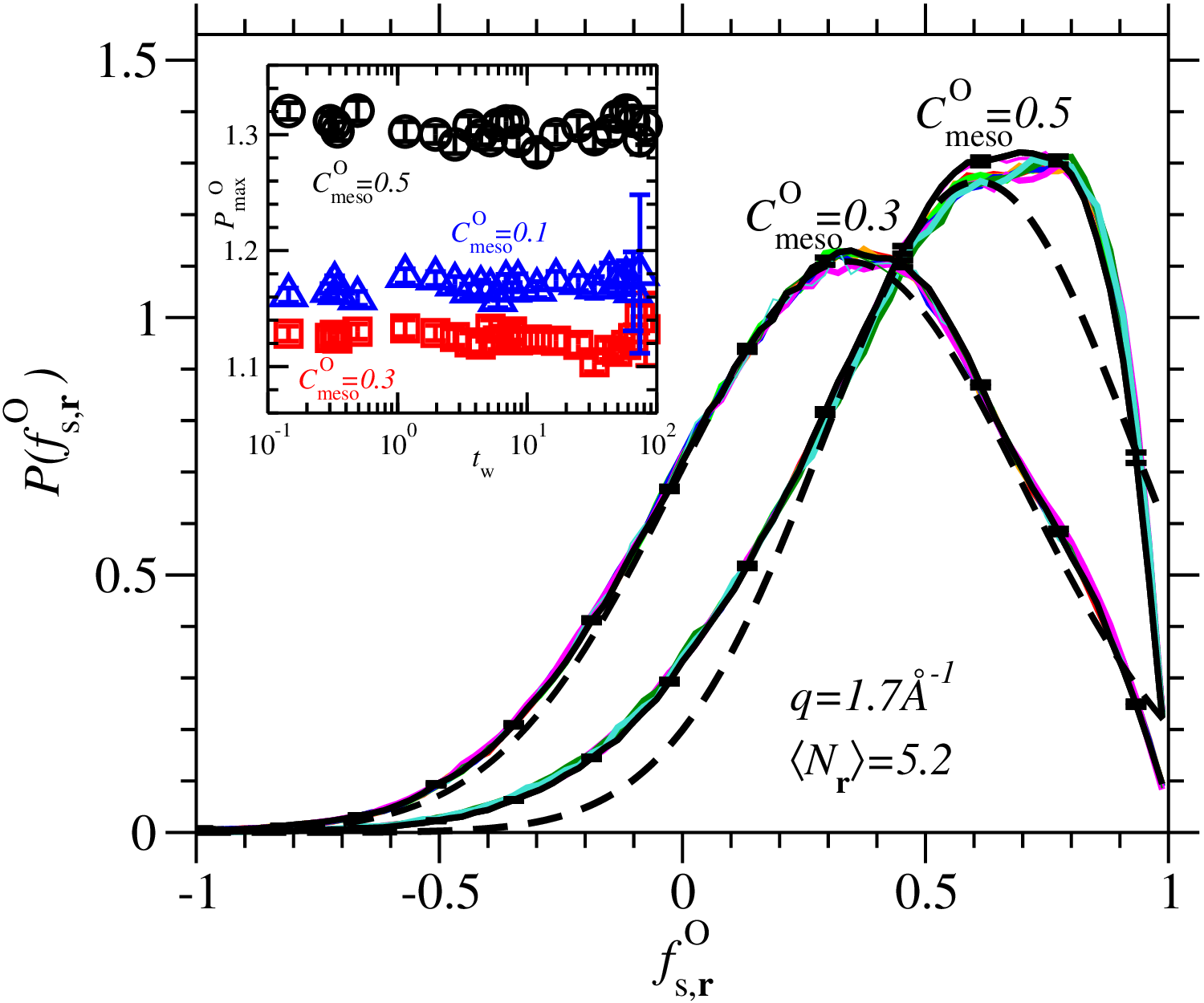}
\caption{\sf (color online) 
 Distribution of the local incoherent intermediate scattering
 function. $P(f_{{\rm s},{\mathbf r}}^{\alpha})$ is an average over
 single simulation run distributions 
$P_c(f_{s,{\mathbf r}}^{\alpha}(t_{\rm w},t^{\alpha}_{\rm meso}(c),q))$
where $t^{\alpha}_{\rm meso}(c)$ is chosen such that $C_{\rm meso}^{\alpha}$
is fixed. The inset shows the maximum value as a function of 
waiting time $t_{\rm w}$.
For comparison with Fig.~\ref{fig:pofcq_q17_Cqfix050_030_O_b2_insetPmax}
we show the case of $q=1.7 \AA^{-1}$ 
and $\langle N_{\mathbf r} \rangle=5.2$
for oxygen atoms, but here we fixed $C_{\rm meso}^{\alpha}$ instead
of $C^{\alpha}$.
}
\label{fig:pofcqcqfix_q17_Cqfix050_030_O_b2_insetPmax}
\end{figure}
%-------------

%\medskip
The resulting average distribution $P(f_{{\rm s},{\mathbf
    r}}^{\alpha})$ is shown in
Fig.~\ref{fig:pofcqcqfix_q17_Cqfix050_030_O_b2_insetPmax} for $q=1.7
\AA^{-1}$ and $\langle N_{\mathbf r} \rangle=5.2$. The evolution of
the maximum $P_{\rm max}^{\rm O}$ with waiting time is shown in the
inset. The comparison with
Fig.~\ref{fig:pofcq_q17_Cqfix050_030_O_b2_insetPmax} confirms that the
scaling is drastically improved by fixing $C_{\rm meso}^{\alpha}$
instead of $C^{\alpha}$.  To test the limits of this improved scaling,
we show in
Fig.~\ref{fig:pofcqcqfix_q17_Cqfix050_030_010_SiO_b4_insetPmax} the
distributions $P(f_{{\rm s},{\mathbf r}})$ for $q=1.7$~\AA$^{-1}$,
$\langle N_{\mathbf r} \rangle = 42$ and $\alpha=$all. This is the
most unfavorable case, with the lowest wavevector we have considered,
the largest coarse graining region, and including both O and Si atoms
(which additionally tests whether the aging clock is the same for both
particle types). Even for this most unfavorable case, we find almost
perfect scaling. We thus conclude that $C_{\rm meso}^{\alpha}$ is the
appropriate scaling quantity.

%-------------
\begin{figure}
\includegraphics[width=3.4in]{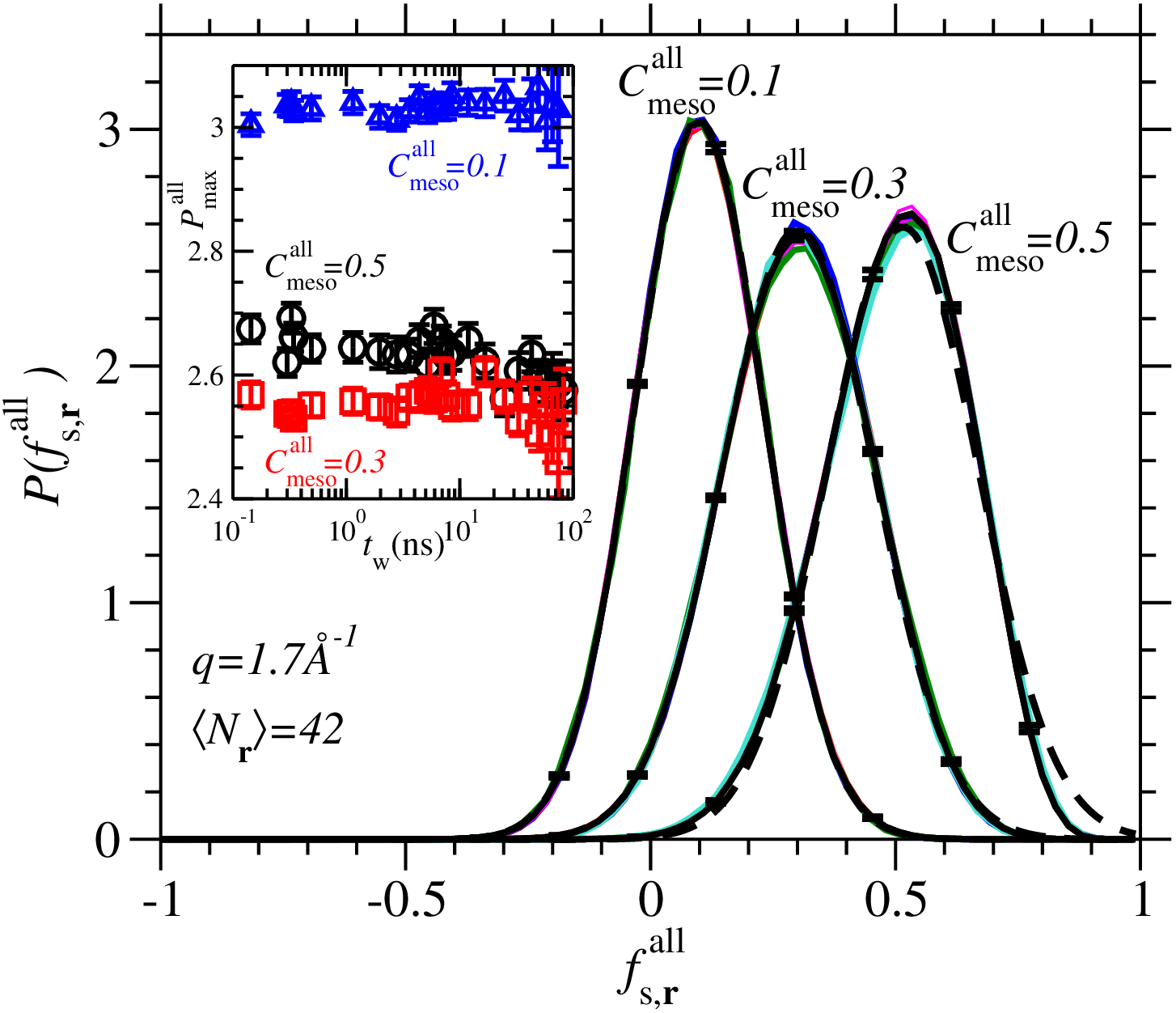}
\caption{\sf (color online) Probability distributions of local
  incoherent intermediate scattering functions, with $\alpha=$all
  (i.e. both O and Si atoms are included), for small $q=1.7 \AA^{-1}$
  and large $\langle N_{\mathbf r} \rangle =42$, for different waiting
  times. The color coding for the waiting times is the same as in
  Fig.~\ref{fig:pofcq_q27_Cqfix050_O_b2}.  Despite this being the most
  unfavorable case for a successful collapse, we find that choosing
  $C_{\rm meso}^{\alpha}$ to be constant results in an almost perfect
  data collapse among results for different waiting times. }
\label{fig:pofcqcqfix_q17_Cqfix050_030_010_SiO_b4_insetPmax}
\end{figure}
%-------------

\section{Conclusions}
\label{sec:conclusions}

In summary, we have performed molecular dynamics simulations of the
strong glass former SiO$_2$ to investigate the scaling of dynamical
heterogeneities in this system. We have quenched the system from an
initial high temperature $T_i$ to a final temperature $T_f$ below the
mode-coupling critical temperature $T_c$, and observed the out of
equilibrium dynamics as a function of the waiting time $t_{\rm w}$,
the time elapsed since the temperature quench.  In particular, we
have investigated the global incoherent intermediate scattering function
$C^{\alpha}(t_{\rm w},t_{\rm w}+t,q)$, the dynamic susceptibility
$\chi_4^{\alpha}(t_{\rm w},t_{\rm w}+t,q)$, and the distribution
$P(f_{{\rm s},{\mathbf r}}^{\alpha}(t_{\rm w},t_{\rm w}+t,q))$ of the
local incoherent intermediate scattering function, where $q$
corresponds to the wave vector magnitude and $\alpha$ specifies the
particle type.

%\medskip
We have found that for sufficiently long waiting times $t_{\rm w}$,
and when probing small enough regions in the system, the dependence on
$(t_{\rm w},t)$ of $\chi_4^{\alpha}(t_{\rm w},t_{\rm w}+t,q)$ and of
$P(f_{{\rm s},{\mathbf r}}^{\alpha}(t_{\rm w},t_{\rm w}+t,q))$ is
governed by $C^{\alpha}(t_{\rm w},t_{\rm w}+t,q)$, up to a $t_{\rm
  w}$-dependent scale factor in $\chi_4$. This is consistent with
predictions for spin glasses and similar to previous results for
fragile glass formers. We thus conclude that the behavior of dynamical
heterogeneity in the aging regime of glassy systems shows a remarkable
degree of universality. A similarity of the behavior for strong and
fragile glass formers had previously been shown for the microscopic
dynamics of single particle jumps \cite{kvlPRL2013}, but here we have
shown that it extends to the scaling of dynamical heterogeneities.

%\medskip
Furthermore we have studied {\em directly\/} the influence of the particle
type $\alpha$ on the dynamics. We have found that $C^{\alpha}(t_{\rm
  w},t_{\rm w}+t,q) = C(z(t_{\rm w},t),q,\alpha)$, where $z(t_{\rm
  w},t)$ is independent of $q$ and $\alpha$. Thus $z(t_{\rm w},t)$
plays the role of a ``common aging clock'' that determines the slow
aging behavior of the relaxation for both Si and O atoms. By combining
this statement with the fact, discussed above, that the aging of
$\chi_4^{\alpha}$ and of $P(f_{{\rm s},{\mathbf r}}^{\alpha})$ is
controlled by $C^{\alpha}$, it follows that the aging of
$\chi_4^{\alpha}$ and of $P(f_{{\rm s},{\mathbf r}}^{\alpha})$ for
both Si and O atoms should be controlled by that same unique aging
clock. Our results show that this prediction is indeed satisfied. In
summary, we have found that both the average and the fluctuations of
the slow aging dynamics are controlled by a unique aging clock, which
is independent of the wavevector $q$ and is the same for O and Si
atoms.

When fluctuations are probed over larger regions, either by taking $q
\le 1.7$~\AA$^{-1}$, or by considering larger coarse graining regions
containing around $40$ particles, new phenomena emerge, presumably due
at least in part to the fact that the probed regions contain more than
one correlation volume. In particular, the scaling of $P(f_{{\rm
    s},{\mathbf r}})$ discussed above no longer holds in its initial
form. It is clear that the probability distributions $P_c(f_{{\rm
    s},{\mathbf r}})$ obtained from the independent runs
$c=1,\cdots,200$ vary dramatically from run to run if the time
interval is kept the same across runs. In other words, there is a
nontrivial distribution of distributions. This is equivalent to the
statement that, in a very large system, a measurement of $P(f_{{\rm
    s},{\mathbf r}})$ over a mesoscopic region containing a few
hundred particles is {\em not\/} self-averaging, and that a new
significant intermediate lengthscale emerges. It is, however,
possible to recover an excellent collapse of probability distributions
for different waiting times $t_{\rm w}$ if instead of averaging
probability distributions $P_c(f_{{\rm s},{\mathbf r}})$ from
different runs (or mesoscopic regions) at constant time interval
$t=t_{\rm fix}$, one averages probability distributions taken at
constant {\em mesoscopic\/} intermediate scattering function $C_{\rm
  meso}^{\alpha}$.

%-------------

\begin{acknowledgments}
We thank A. Parsaeian for preliminary work. KVL and HEC thank
A. Zippelius and the Institute of Theoretical Physics, University of
G{\"{o}}ttingen, for hospitality and financial support. CHG was
supported by NSF REU Grant PHY-1156964.  This work was supported in
part by the Deutsche Forschungsgemeinschaft via SFB 602
and FOR1394, by DOE under grant DE-FG02-06ER46300, and by Ohio
University. Numerical simulations were carried out at Bucknell 
University and Ohio University. Part of this work was performed at the Aspen Center for
Physics, which is supported by National Science Foundation grant
PHY-1066293. 
\end{acknowledgments}

\bibliography{scalingSiO2_v15}% Produces the bibliography via BibTeX

\end{document}